\documentclass[%
reprint,
superscriptaddress,
amsmath,amssymb,
aps,
prb,
floatfix,
]{revtex4-2}

\usepackage{graphicx}
\usepackage{dcolumn}
\usepackage{bm}
\usepackage{xcolor}
\usepackage{xr-hyper}
\usepackage{xr}

\usepackage{amsmath}
\usepackage{dirtytalk}
\usepackage{tabularx}
\usepackage{multirow}
\usepackage{makecell}
\usepackage[version=4]{mhchem}

\usepackage{hyperref}
\begin{document}

\preprint{APS/123-QED}

\title[PRB]{Electronic structure, band offset, and interface electron population of the \texorpdfstring{\ce{LaInO3}}{LaInO3}/\texorpdfstring{\ce{BaSnO3}}{BaSnO3} system}

\author{G.~Hoffmann}
\affiliation{Paul-Drude-Institut für Festkörperelektronik, Leibniz-Institut im Forschungsverbund Berlin e.\,V., Hausvogteiplatz 5--7, 10117 Berlin, Germany.}

\author{A.~A.~Riaz}
\affiliation{Department of Chemistry, University College London, 20 Gordon Street, London, WC1H~0AJ, United Kingdom.}

\author{C.~Kalha}
\affiliation{Department of Chemistry, University College London, 20 Gordon Street, London, WC1H~0AJ, United Kingdom.}

\author{A.~Gloskovskii}
\author{C.~Schlueter}
\affiliation{Deutsches Elektronen-Synchrotron DESY, Notkestrasse 85, 22607 Hamburg, Germany.}

\author{O.~Brandt}
\affiliation{Paul-Drude-Institut für Festkörperelektronik, Leibniz-Institut im Forschungsverbund Berlin e.\,V., Hausvogteiplatz 5--7, 10117 Berlin, Germany.}

\author{W.~Aggoune}
\author{C.~Draxl}
\affiliation{Institut f{\"u}r Physik and CSMB, Humboldt-Universit{\"a}t zu Berlin, 12489 Berlin, Germany.}

\author{O.~Bierwagen}
\affiliation{Paul-Drude-Institut für Festkörperelektronik, Leibniz-Institut im Forschungsverbund Berlin e.\,V., Hausvogteiplatz 5--7, 10117 Berlin, Germany.}

\author{A.~Regoutz}
 \email{anna.regoutz@chem.ox.ac.uk}
 \affiliation{Department of Chemistry, University of Oxford, Inorganic Chemistry Laboratory, South Parks Road, Oxford, OX1~3QR, United Kingdom.}
\affiliation{Department of Chemistry, University College London, 20 Gordon Street, London, WC1H~0AJ, United Kingdom.}

\date{\today}

\begin{abstract}

Perovskite oxides and their heterostructures exhibit a wide range of functional properties. Among these materials, \ce{BaSnO3}/\ce{LaInO3} heterostructures form high-mobility two-dimensional electron gases (2DEGs) at their interfaces.
In particular, room-temperature electron mobilities exceeding 100~cm$^2$/Vs were enabled by recent advances in thin-film growth. This work presents a combined experimental and theoretical study of the electronic structure of \ce{BaSnO3}, \ce{LaInO3}, and \ce{BaSnO3}/\ce{LaInO3} heterostructures with
varying \ce{LaInO3} overlayer thicknesses. Soft and hard X-ray photoelectron spectroscopy (SXPS and HAXPES) measurements are combined with densities of states (DOS) derived from hybrid density functional theory (DFT) calculations. The analysis of core, semi-core, and valence states allows to arrive at a comprehensive understanding of the chemical bonding and electronic structure in the parent oxides as well as the formed heterostructures. For the  \ce{BaSnO3}/\ce{LaInO3} heterostructure, the band offset and population of 2DEG states at the interface is directly probed using HAXPES.

\end{abstract}

\maketitle

\section{Introduction}

Ternary oxides, particularly those exhibiting the perovskite structure, are of considerable interest due to their innate ability to exhibit diverse optical, electronic, and magnetic characteristics. \ce{BaSnO3} and \ce{LaInO3} are two such oxides that have garnered attention, not only by themselves, but particularly as their combination enables the formation of a high-mobility, two-dimensional electron gas (2DEG) at the epitaxially formed interface due to a polar discontinuity.~\cite{Aggoune2021} By using specifically optimised growth conditions to prepare \ce{BaSnO3}/\ce{LaInO3} interfaces, we have recently demonstrated the achievement of room temperature (RT) electron mobility values above 100~cm$^2$/Vs at a given carrier concentration in the mid 10$^{13}$ cm$^{-2}$.~\cite{Hoffmann2024} Furthermore, values of more than 2000~cm$^2$/Vs at carrier concentrations in the high 10$^{13}$~cm$^{-2}$ have been reported at cryogenic temperatures by electric-double-layer gating.~\cite{Lee2025} Although the latter approach is limited to laboratory applications, it showcases the incredible potential of these heterostructures in the field of oxides for both RT device applications and low-temperature fundamental research. \par

Given the tremendous promise of the two oxides, both experimental and theoretical efforts have been made to understand their electronic structure in detail. The first reports on the electronic structure of \ce{BaSnO3} in the 1990s were in relation to the presence (or absence) of superconductivity in complex oxides.~\cite{Singh1991} Since then, it has been explored in great detail using theoretical approaches, such as density functional theory (DFT) and many-body perturbation theory,~\cite{Moreira2012,Duarte2016,Weston2018,Rowberg2020,Aggoune2022} including studies on the influence of La, ~\cite{Sallis2013,Lebens-Higgins2016} and Sb doping.~\cite{Kim2014} Spectroscopy approaches have been used in parallel to probe the electronic structure experimentally, including photoelectron spectroscopy (PES),~\cite{Sallis2013,Lebens-Higgins2016,Joo2017,NonoTchiomo2022} and electron energy-loss
spectroscopy (EELS).~\cite{Yun2018,Aggoune2022} \ce{LaInO3} has received much less attention as a stand-alone material, with only a small number of theoretical works dedicated to it.~\cite{Erkişi2016,Galazka2021,Aggoune+LIO} Two studies by some of the authors of the present work explored the electronic structure of \ce{LaInO3} and other lanthanide indium perovskites by combining first-principles theory with optical absorption measurements ~\cite{Aggoune+LIO} and PES,~\cite{Hartley2021} respectively. \par

By extension from the isolated oxides, the combination of \ce{BaSnO3} and \ce{LaInO3} has been at the centre of work by several groups. Notably, the group led by Kookrin Char has developed field-effect transistors (FETs) based on \ce{BaSnO3} as the channel layer and \ce{LaInO3} as the gate oxide since the mid-2010s.~\cite{Kim2015,Kim2019,Kim2023,Lee2025} Theoretically, K.~Krishnaswamy and coworkers used hybrid DFT to explore \ce{BaSnO3} in combination with several oxides, including \ce{LaInO3}.~\cite{Krishnaswamy2016,Krishnaswamy2017} Recent studies by these authors form the foundation of the present work, including growth, experimental characterisation, and theory.~\cite{Hoffmann2024,Zupancic2024,Aggoune2021,Fang2023} 

This work reports a comprehensive exploration of the electronic structure of \ce{BaSnO3}, \ce{LaInO3}, and a series of \ce{BaSnO3}/\ce{LaInO3} heterostructures with varying \ce{LaInO3} overlayer thicknesses, combining both hybrid DFT and soft and hard X-ray PES (SXPS and HAXPES). In particular, semi-core and valence spectra from experiment are compared to photoionisation cross-section-corrected projected densities of states (PDOS) to explore the chemical bonding and electronic structure of the two parent oxides, as well as heterostructures based on their combination. \par

\section{Experimental Methods}

\begin{figure}
        \centering
        \includegraphics[width=1\linewidth]{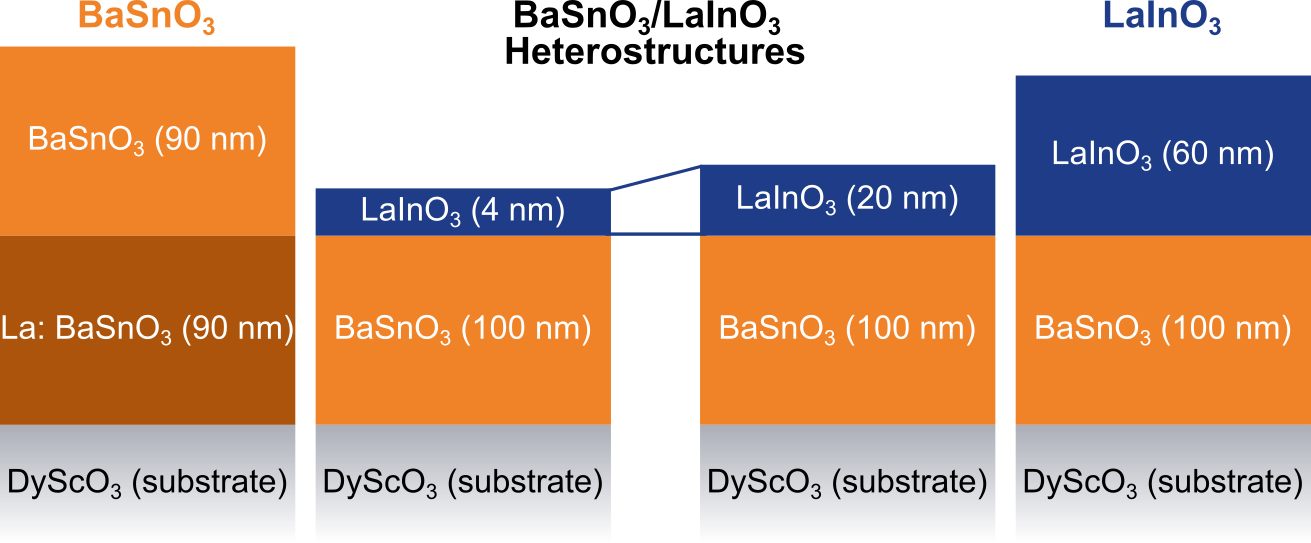}
        \caption{Schematic of the design of \ce{BaSnO3}, \ce{LaInO3}, and \ce{BaSnO3}/\ce{LaInO3}  heterostructure samples. The heterostructures comprise \ce{LaInO3} layers with thicknesses ranging from 4 to 20~nm. All samples were grown on \ce{DyScO3} (110) substrates.}
        \label{fig:schematic}
    \end{figure}

All samples were grown using plasma-assisted molecular beam epitaxy (PA-MBE) in a custom-made UHV chamber on \ce{DyScO3}(110) substrates. Details on the growth parameters and substrate preparation can be found elsewhere \cite{Hoffmann2024}. Figure~\ref{fig:schematic} provides a schematic overview of the film stacks and thicknesses prepared. For the heterostructures, 100~nm \ce{BaSnO3}(100) was grown on \ce{DyScO3}(110) substrates, followed by a \ce{LaInO3} layer with varying thicknesses of 4, 6, 8, 14, 16, and 20~nm. Note that \ce{DyScO3} and \ce{LaInO3} are orthorhombic, but allow for definition of a pseudo cubic unit cell with pseudo cubic $\{100\}_\text{pc}$ lattice planes that are almost parallel to the \ce{BaSnO3}(100) and \ce{DyScO3}(110) orientation).~\cite{Hoffmann2023,Zupancic2020} As a consequence, the derived out-of-plane lattice constants c$_{pc}$~=~c/2 of DyScO$_3$ and LaInO$_3$ are c$_{pc}$~=~0.39522~nm,~\cite{Schmidbauer2012} and c$_{pc}$~=~0.41107~nm, \cite{Galazka2021} respectively, while for BaSnO$_3$ c~=~0.4116~nm.~\cite{Galazka2017}

The SnO$_2$/LaO interface termination of the \ce{BaSnO3}/\ce{LaInO3} heterostructure (essential for 2-DEG formation) was engineered using a shutter-controlled growth sequence across the interface to realise an interfacial 2DEG in the \ce{BaSnO3} layer, as detailed in our previous work.~\cite{Hoffmann2024} 

Since undoped \ce{BaSnO3} and \ce{LaInO3} are wide band gap semiconductors with indirect and direct band gaps of 3.0 and 4.35~eV,~\cite{Galazka2017,Galazka2021} respectively, it is not easily possible to measure reference bulk samples using PES due to the electric charging of the samples. Consequently, heterostructures were grown as described below. For the \ce{BaSnO3} reference sample, 90~nm of La-doped \ce{BaSnO3} was grown on a \ce{DyScO3}(110) substrate, followed by 120~nm undoped \ce{BaSnO3}, with the La:\ce{BaSnO3} layer helping to compensate any charging. For the \ce{LaInO3} reference, a \ce{BaSnO3}/\ce{LaInO3} heterostructure was grown in the same manner as described above, but with a 60~nm thick \ce{LaInO3} layer. While the interfacial 2DEG compensates electric charging, the thickness of the top layer ensures spectra to only arise from the undoped \ce{LaInO3}.

After growth, the structural quality of the samples was determined using X-ray diffraction (XRD) and atomic force microscopy (AFM). For analysis of the surface morphology, a Bruker Dimension Edge atomic force microscope was used in peak force tapping mode. For analysis of the crystal quality, a X'pert pro MRF diffractometer from Philips PANalytical with Cu~K$\alpha$ radiation of $\lambda = 0.154$~nm was used. While the wide-angle $2\Theta - \omega$ scan did not show any additional phases, the scan of the $\omega$ rocking curve provides further insights into crystal quality.\newline

For identification of the 2DEG, capacitance-voltage (CV) measurements were performed (see Figure~S1 in the Supplementary Information). For electrical characterisation, Ti/Au contacts with thicknesses of 20/100~nm were applied to the corners and edges of the samples using a shadow mask by an e-beam evaporator, Pfeiffer Classic 500. 

Hard X-Ray Photoelectron Spectroscopy (HAXPES) was performed at beamline P22 at PETRAIII, German Electron Synchrotron DESY in Hamburg, Germany.~\cite{Schlueter2019} Photon energies of 3.4, 6.0, and 8.0~keV were used. A Si(311) double-crystal monochromator was used to achieve the two higher photon energies of 6.0 and 8.0~keV, whilst a double channel cut post-monochromator employing a Si(111) and a Si(220) channel-cut crystal pair was used to select 3.4~keV. The HAXPES endstation uses a Phoibos 225HV analyser (SPECS, Berlin, Germany), which was used in the small area lens mode with a slit size of 3~mm. Spectra were collected using a pass energy of 30~eV. The total energy resolution for the photon energies in this setup was determined based on the 16/84\% Fermi edge ($E_F$) width of a scraped, polycrystalline gold foil. The resulting total energy resolutions were 215 and 249~meV for 3.4 and 6.0~keV, respectively. All experiments were conducted in near grazing incidence geometry (below 5$^\circ$). All samples were mounted using conductive carbon tape (Nisshin-EM) on the substrate as well as the sample edge and surface. For the reference \ce{BaSnO3} and \ce{LaInO3} samples, comprehensive datasets were only collected at 6.0~keV due to time constraints at the synchrotron.
To complement the synchrotron-based HAXPES experiments, laboratory-based SXPS was collected on a Thermo K-Alpha+ spectrometer. The instrument operates with a monochromated Al~K$\alpha$ photon source ($h\nu$ = 1.4867~keV). It uses a 180$^\circ$ double-focusing hemispherical analyser, a two-dimensional detector, and operates at a base pressure of 2$\times$10\textsuperscript{-9}~mbar. A 400~$\mu$m spot size was used for all measurements, achieved by operating the X-ray anode at an emission current of 6~mA and a cathode voltage of 12~kV. A flood gun with an emission current of 100~$\mu$A was used to achieve the desired level of charge compensation. The total energy resolution of the spectrometer was determined to be 440~meV. Survey,  core level, and valence spectra were collected with pass energies of 200, 20, and 15~eV, respectively.

\section{Computational Methods}
All calculations were performed using FHI-aims,~\cite{FHI-aims} an all-electron full-potential package employing numerical atom-centered orbitals. For optimizing the bulk structures a \textit{tight} setting was used. The generalized gradient approximation (GGA) in the PBEsol parametrization was considered for the exchange correlation effects.~\cite{PBEsol+08prl} Brillouin zone integrations were performed on a 8$\times$8$\times$8$~\textbf{k}$-grid for BaSnO$_3$ and a 6$\times$6$\times$4$~\textbf{k}$-grid for LaInO$_3$. Internal coordinates and lattice parameters were optimized until the residual forces on each atom were less than 0.001 eV \AA$^{-1}$.

Electronic properties were calculated using the hybrid functional HSE06~\cite{HSE+06jcp}, incorporating 33\% (BaSnO$_3$) and 30\% (LaInO$_3$) of the Hartree-Fock exchange, giving band gaps of 3.1 eV and 4.6 eV, respectively, in close agreement with the experiment.~\cite{Aggoune+LIO,Aggoune2022} Spin-orbit coupling was included as it affects the considered semi-core states.~\cite{Aggoune+LIO,Aggoune2022} This allows for a quantitative comparison with experiment. Projected densities of states were computed applying a Gaussian broadening of 0.01 eV. 

\subsection{Comparison of Theory and Experiment}\label{X_section_method}

To directly compare the theoretically-derived PDOS with the experimental semi-core level (SCL) and valence band (VB) spectra, the PDOS have to be (i) broadened, (ii) photoionisation cross section corrected, and (iii) aligned to a common energy reference. To address (i), the PDOS were broadened using Gaussian broadening corresponding to 100$\%$ of the experimental $E_F$(Au), i.e. 0.323~eV for measurements at 6.0 keV and 3.4~keV, and 0.647~eV for 1.5 keV. In addition, a Lorentzian broadening of 0.2~eV was applied to account for average lifetime broadening effects. The PDOS were aligned such that the VB onset (50$\%$ value) and characteristic peaks in the VB approximately match the experimental data. For SCL spectra, the PDOS was aligned to the lowest binding energy feature.

Photoionisation cross-section corrections were applied to the PDOS to scale the contributions from specific elements and orbitals, based on theoretical cross sections from Scofield.~\cite{Scofield1973,Kalha2020} To allow for the correction of La~5\textit{d} and Ba~5\textit{d} projected states, which are not occupied in the ground state, Ce~5\textit{d} cross-section values were employed, as Ce is the next element to have this orbital at least partially occupied, meaning that theoretical cross section values are available. Finally, the so-derived theoretical and corresponding experimental spectra were normalised to their respective areas for the VB and to the height of the lowest binding energy feature for the SCL spectra, respectively.

\section{Results and Discussion}

\begin{figure}
    \centering
    \includegraphics[width=0.8\linewidth]{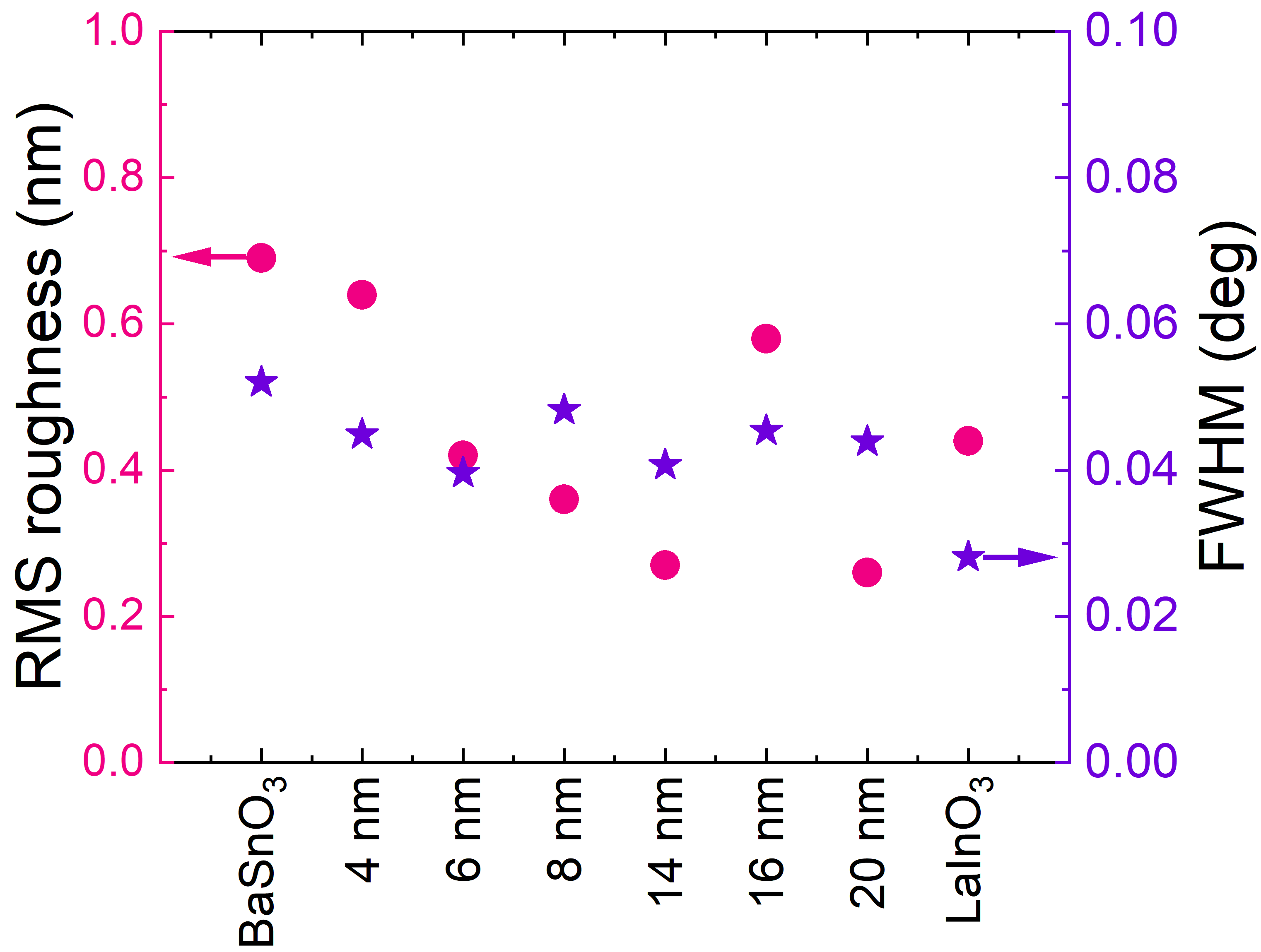}
    \caption{Overview of AFM surface root mean square (RMS) roughness (left axis) and full width at half maximum (FWHM) values of the BSO 200 rocking curve from XRD (right axis) for all samples investigated in this study, including the \ce{BaSnO3} and \ce{LaInO3} reference samples as well as the \ce{BaSnO3}/\ce{LaInO3} heterostructures with \ce{LaInO3} layer thicknesses of 4 to 20~nm.}    
    \label{fig:afm_xrd_overview}
\end{figure}

The crystal quality of the grown samples was investigated using atomic force microscopy (AFM) and X-ray diffraction (XRD) (see Figures~S2 and S3 in the Supplementary Information for the AFM images and XRD 2$\Theta$-$\omega$ scans). Fig.~\ref{fig:afm_xrd_overview} shows root mean square (RMS) roughness values derived from AFM as well as full width at half maximum (FWHM) values of the \ce{BaSnO3} 200 rocking curve from XRD for all samples investigated in this study. The RMS values vary in the range from 0.26 to 0.65~nm, typical for oxide thin film systems prepared by PA-MBE. In parallel, the FWHM values also show a typical variation of between 0.028 and 0.052$^{\circ}$.\\

Survey spectra for all samples were collected to enable specification of high-resolution spectral collection and are shown in Figures~S4 and S5 in the Supplementary Information for the \ce{BaSnO3} and \ce{LaInO3} references, and the \ce{BaSnO3}/\ce{LaInO3} heterostructures, respectively. Survey spectra show all expected core lines and their relative intensities change as expected from the layer design and thickness. In addition, and as explored in previous work,~\cite{Zupancic2024} Ba segregation to the surface of \ce{LaInO3} is observed, particularly in the surface-sensitive measurements at 1.5~keV, with Ba core lines appearing in the spectra. 

Based on the TPP-2M model as implemented in the QUASES software package,~\cite{Shinotsuka2015} the inelastic mean free paths (IMFPs) for \ce{BaSnO3} and \ce{LaInO3} are comparable, with maximum IMFPs at 6~keV being 75.8 and 77.6~\AA, respectively. The parameters used in the model and values for all photon energies employed are given in Tables~S1 and S2, respectively, in the Supplementary Information.  

For each sample, key core, as well as semi-core and valence spectra, were collected and are discussed in the following sections. 

\subsection{\texorpdfstring{\ce{BaSnO3}}{BaSnO3}}

\subsubsection{Core States}

Key core state spectra for \ce{BaSnO3} collected at 6.0~keV are shown in Figure~\ref{fig:BSO_LIO_CLs} (the corresponding O~1s spectra are included in Figure~S6 in the Supplementary Information), and their spectral shapes and binding energy positions are commensurate with an extended oxide. The Ba~3\textit{d}\textsubscript{5/2} line is at a $E_B$ position of 779.4~eV (see Figure~\ref{fig:BSO_LIO_CLs}(a)) and the Sn~3\textit{d}\textsubscript{5/2} line at 486.5~eV (see Figure~\ref{fig:BSO_LIO_CLs}(b)). The Sn~3\textit{d} core level shows a spin-orbit splitting (SOS) of 8.4~eV. The SOS of the Ba~3d core level is more than 15~eV and, therefore, only the 3\textit{d}\textsubscript{5/2} line was measured to allow for efficient use of synchrotron time. The HAXPES data provide high-quality reference spectra, as they are not influenced by surface states (e.g., adventitious C\textsuperscript{0}, hydroxyl species, or surface-segregated Ba). Surface-sensitive spectra were collected at 1.5~keV and are presented in Figure~S7 in the Supplementary Information for completeness. Overall, the sharp core-level features and absence of detectable additional chemical components are consistent with a chemically homogeneous, phase-pure \ce{BaSnO3} film, with no evidence for significant secondary phases or strongly distinct defect-related environments.

    \begin{figure}
        \centering
        \includegraphics[width=1.0\linewidth]{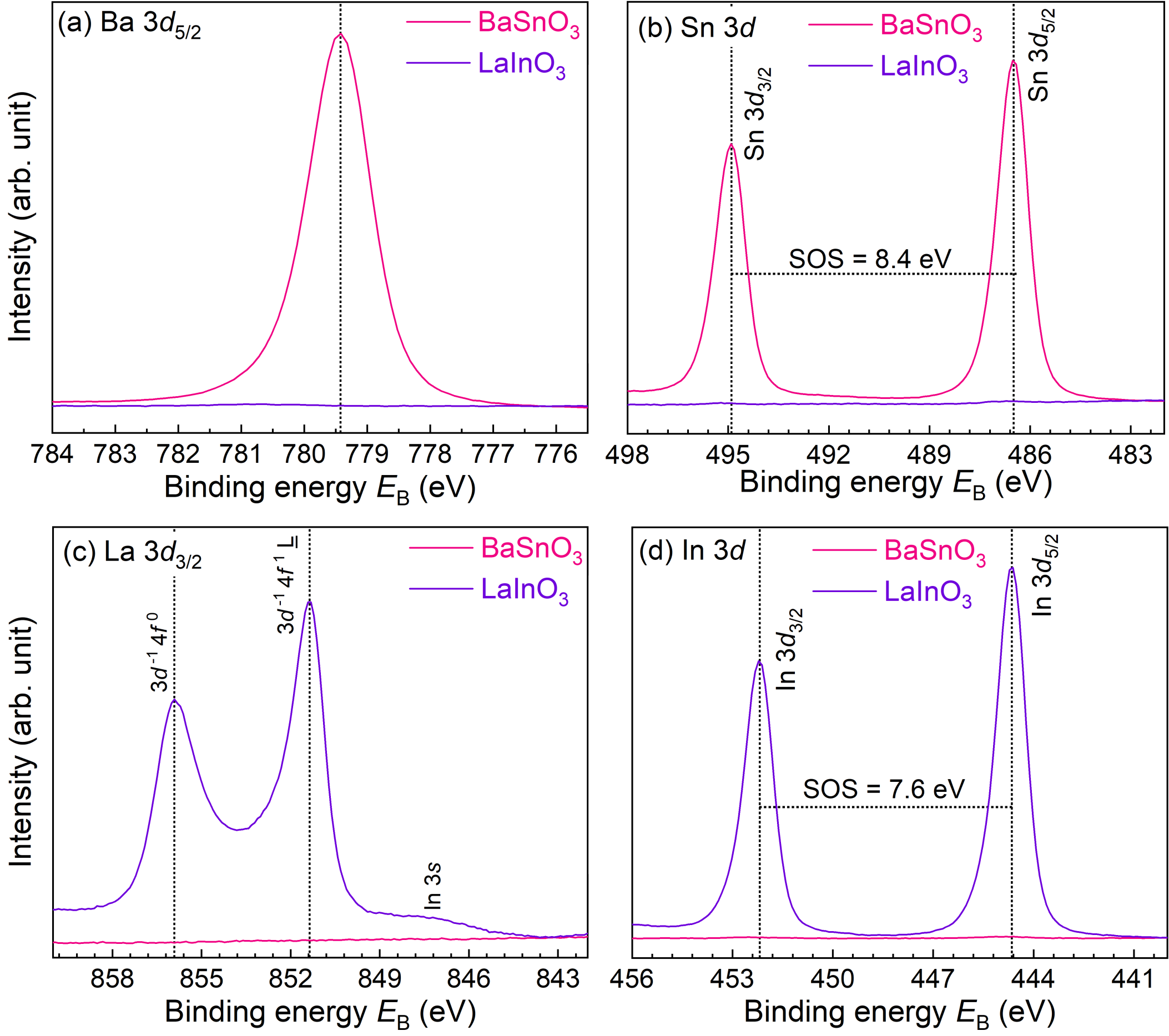}
        \caption{Key core state spectra of \ce{BaSnO3} and \ce{LaInO3} collected at a photon energy of 6.0~keV, including (a) Ba~3\textit{d}\textsubscript{5/2}, (b) Sn~3\textit{d}, (c) La~3\textit{d}\textsubscript{3/2}, and (d) In~3\textit{d}. For the Sn and In~3\textit{d} doublets, the value of the experimental spin-orbit splitting (SOS) is included.}
        \label{fig:BSO_LIO_CLs}
    \end{figure}

\subsubsection{Semi-core States}

\begin{figure}
        \centering
        \includegraphics[width=0.7\linewidth]{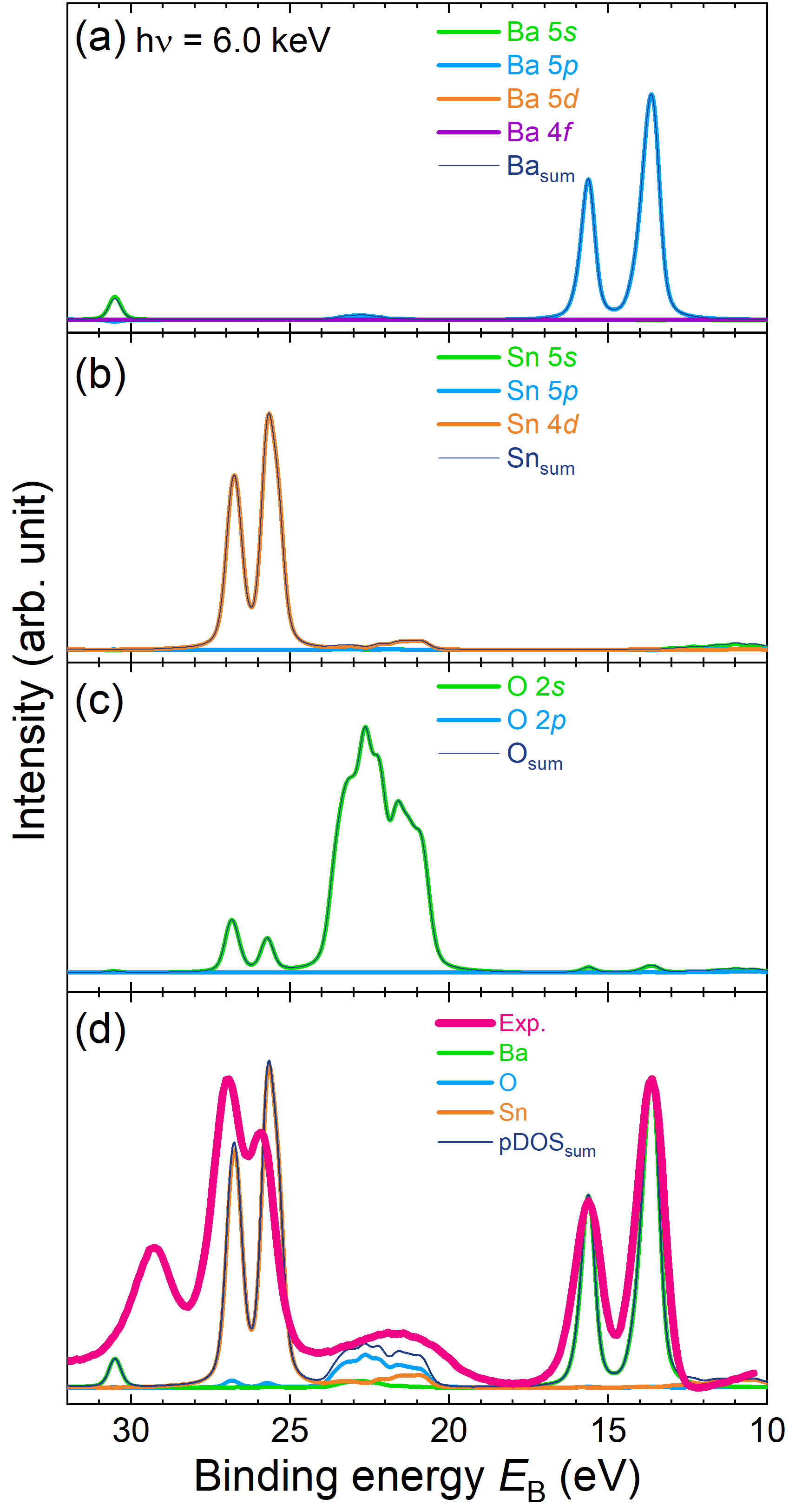}
        \caption{Comparison of the projected densities of state (PDOS) for the semi-core states from DFT ((a)-(c)) with experiment of \ce{BaSnO3}. (a)-(c) Element- and orbital-specific one-electron photoionisation cross section corrected projected densities of state (PDOS) from DFT, and (d) total element-specific one-electron photoionisation cross section corrected PDOS and HAXPES data (Exp.) collected at a photon energy of 6.0~keV. In (d), experiment and theory were normalised to the feature height of the lowest binding energy $E_B$ semi-core state (Ba~5\textit{p}\textsubscript{3/2}), respectively.}
        \label{fig:BSO_SCL_6}
    \end{figure}

Beyond the deeper core levels discussed so far, which confirm the presence of an oxide based on line shape and binding energy position, the semi- or shallow-core states shown in Figure~\ref{fig:BSO_SCL_6} provide further information on the chemical bonding present. Although there are some limitations in the accuracy of the calculated energies for these states, a comparison of the experimental spectrum with the calculated PDOS enables commentary on the level and nature of hybridisation of these states. The experimental spectra show clear doublet features for the Ba~5\textit{p} states at 13.6 and 15.6~eV (resulting in a SOS of 2.0~eV) and for the Sn~4\textit{d} states at 25.9 and 27.0~eV (SOS = 1.1~eV). By comparison with the theoretical spectra, it is clear that the Ba~5\textit{p} states show little hybridisation, whilst the Sn~4\textit{d} states show some mixing with the O~2\textit{p} states. Hybrid functionals, such as HSE06 used in the present work substantially improve band-edge positions compared to e.g., GGA, but still underestimate the binding energy of localised semicore \textit{d} states. We have previously explored this in further detail for the analogous Ga~3\textit{d} states in \ce{Ga2O3}.~\cite{Swallow2020} The underestimation arises due to residual \textit{p}-\textit{d} repulsion and missing quasiparticle relaxation, explaining the small discrepancy observed here for the Sn~4\textit{d} levels in \ce{BaSnO3}. The mixing of the Sn~4\textit{d} and O~2\textit{p} states, coupled with the vicinity to the Ba~5\textit{s} states at 29.3~eV, results in the Sn~4\textit{d}\textsubscript{3/2} peak appearing to have greater intensity than the 4\textit{d}\textsubscript{5/2} peak in the experimental data. The O~2\textit{s} states are considerably mixed with Ba~5\textit{p} and Sn~4\textit{d} states, resulting in a broad feature centred around 21.6~eV. 

\subsubsection{Valence States}
\begin{figure}
        \centering
        \includegraphics[width=0.7\linewidth]{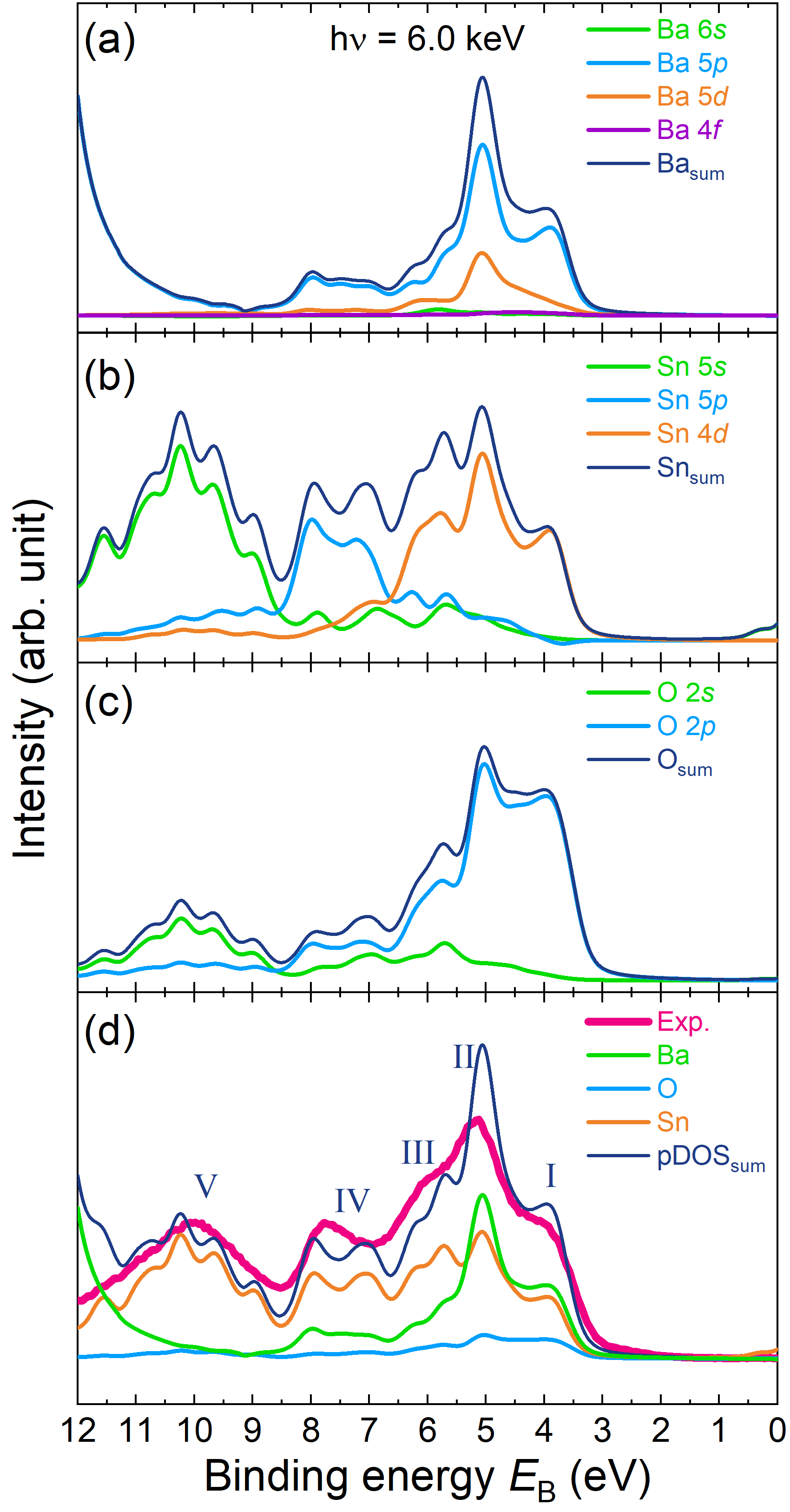}
        \caption{Valence spectra of \ce{BaSnO3} showing (a)-(c) element- and orbital-specific one-electron photoionisation cross section corrected projected densities of state (PDOS) from DFT, and (d) total element-specific one-electron photoionisation cross section corrected PDOS and HAXPES data (Exp.) collected at a photon energy of 6.0~keV. In (d), experiment and theory were normalised to the area of the VB spectra and PDOS between 0 and 12~eV, respectively.}
        \label{fig:BSO_VB_6}
    \end{figure}

Figure~\ref{fig:BSO_VB_6} shows the photoionisation valence band (VB) spectra of the \ce{BaSnO3} reference at 6~keV with five distinct features (I - V) identifiable. At this photon energy the oxygen contribution, which intrinsically dominates the top of the VB, is comparatively small, with Ba and Sn states dominating the VB due to differences in the photoionisation cross sections. At the top of the VB, feature I at 3.95~eV arises from almost equivalent contributions of Ba~5\textit{p} and Sn~4\textit{d} states. In feature II at 5.1~eV, Ba~5\textit{d} states also contribute. In features III-V, Sn states become most dominant with Sn~4\textit{d} (feature III at $\approx$~6.0~eV), Sn~5\textit{p} (feature IV with the higher $E_B$ identifiable peak at 7.8~eV in the experiment), and Sn~5\textit{s} (feature V at 10.0~eV) states being most significant. O~2\textit{p} states contribute mainly at the top of the VB, whilst O~2\textit{s} states are most strongly hybridised with the Sn~5\textit{s} states. These observations are in very good agreement with a previous DFT-HAXPES comparison.~\cite{Sallis2013}

\subsection{\texorpdfstring{\ce{LaInO3}}{LaInO3}}

\subsubsection{Core States}

Core state spectra for \ce{LaInO3} are again commensurate with an extended oxide. The La~3\textit{d}\textsubscript{3/2} (see Figure~\ref{fig:BSO_LIO_CLs}(c)) shows the typical splitting arising from charge-transfer and final state screening.~\cite{Burroughs1976,Kotani1992,Suzuki1998} Briefly, the lower $E_B$ peak at 851.4~eV is associated with a screened final state (3\textit{d}\textsuperscript{-1} 4\textit{f}\textsuperscript{1} \underline{L}) where charge transfer from the ligand, L, valence states (from the oxygen 2\textit{p}) into the 4\textit{f} level occurs which is induced by the core hole potential. The higher $E_B$ component at 855.9~eV is associated with an unscreened (3\textit{d}\textsuperscript{-1} 4\textit{f}\textsuperscript{0}) final state. The 3\textit{d}\textsubscript{3/2} core level was favoured over the 3\textit{d}\textsubscript{5/2} as the latter overlaps with the In~3\textit{s}. The In~3\textit{d} core level has a 3\textit{d}\textsubscript{5/2} $E_B$ of 444.7~eV and a SOS of 7.6~eV. \par

\subsubsection{Semi-core States}

\begin{figure}
        \centering
        \includegraphics[width=0.7\linewidth]{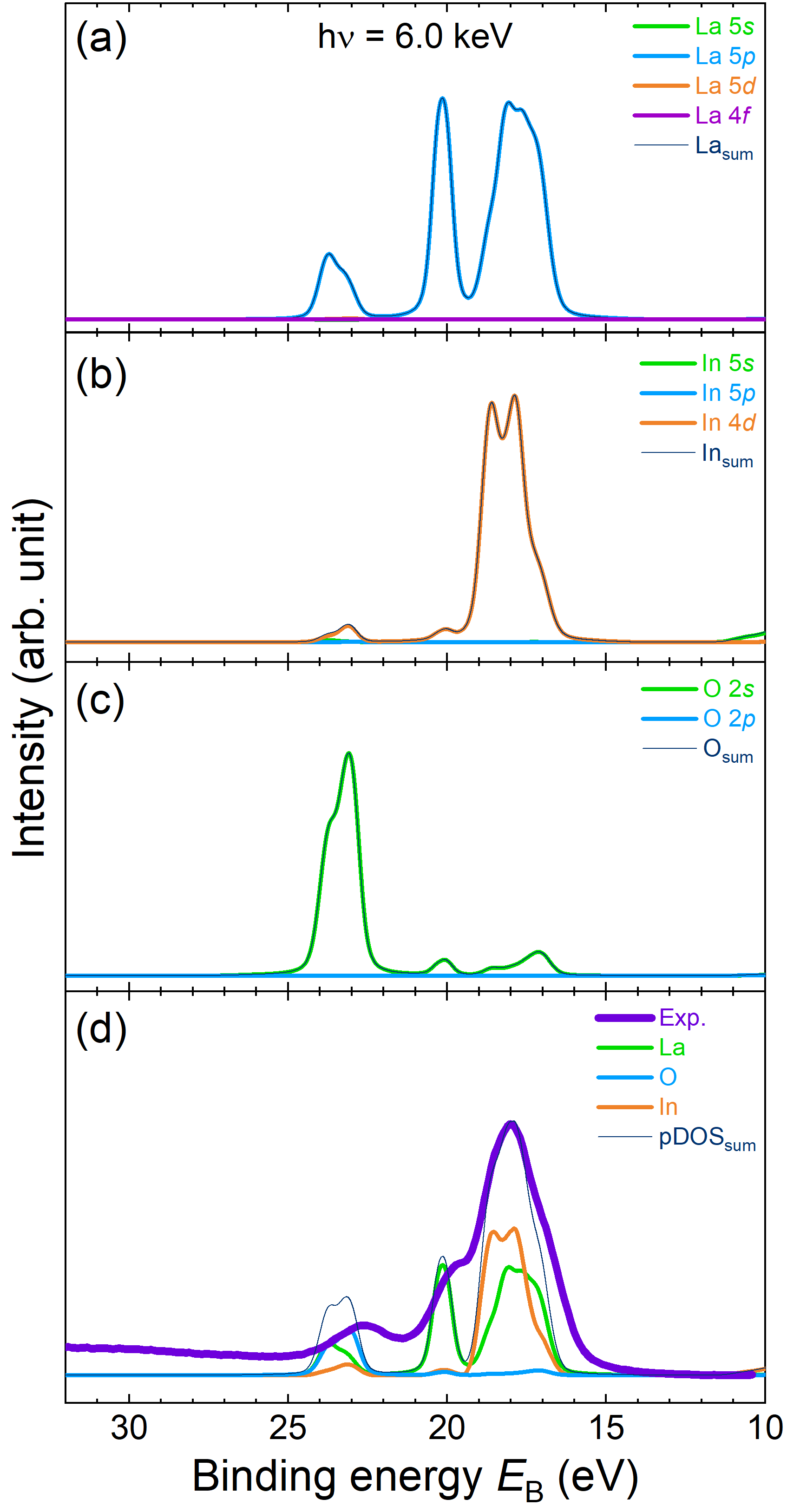}
        \caption{Semi-core state spectra of \ce{LaInO3} showing (a)-(c) element- and orbital-specific one-electron photoionisation cross section corrected projected densities of state (PDOS) from DFT, and (d) total element-specific one-electron photoionisation cross section corrected PDOS and HAXPES data (Exp.) collected at a photon energy of 6.0~keV. In (d), experiment and theory were normalised to the feature height of the highest intensity feature at 18~eV, respectively.}
        \label{fig:LIO_SCL_6}
    \end{figure}

As for \ce{BaSnO3}, the semi-core states provide further insights into the chemical bonding, and specifically hybridisation, present in \ce{LaInO3}, as shown in Figure~\ref{fig:LIO_SCL_6}. Whilst the Ba and Sn states in \ce{BaSnO3} showed only minor mixing and are separated by over 10~eV in $E_B$, there is extensive overlap of La~5\textit{p} and In~4\textit{d} states between 16 and 19~eV, as their isolated atomic orbitals naturally share nearly identical intrinsic binding energies of about 17~eV. A distinct higher $E_B$ La~5\textit{p}-dominated feature is observed at 20~eV. The O~2\textit{s} states have a much narrower energy width (between 22-23~eV in the experiment) compared to \ce{BaSnO3}. The comparatively narrow O~2\textit{s} feature may be attributed to its greater energetic separation from the largely overlapping La~5\textit{p} and In~4\textit{d} states which limits cation--O~2\textit{s} hybridisation relative to \ce{BaSnO3}, where the O~2\textit{s} spectral weight is distributed through interactions with both Ba~5\textit{p} and Sn~4\textit{d} states.

\subsubsection{Valence States}

The top of the VB in \ce{LaInO3}, feature I at 4.2~eV in the experiment, has contributions from La~5\textit{p} and In~4\textit{d} states. Below this, the contributions to  feature II at 5.2~eV are comparable, but with an increase from La~5\textit{d} states. For both features, the La states outweigh the In contributions. This changes somewhat in feature III and IV, where the increase in In~5\textit{p} states leads to a relatively greater contribution from In states overall. Finally, feature V at 8.7~eV is dominated by In~5s states with only minor hybridisation with O and La states. These observations are in good agreement with previous work on the lanthanide indium oxides.~\cite{Hartley2021} Similarly to \ce{BaSnO3}, the VBM in \ce{LaInO3} is mainly dominated by O~2\textit{p} states,~\cite{Aggoune2021} but due to photoionisation cross section effects at 6~keV the La and In contributions dominate.

    \begin{figure}
        \centering
        \includegraphics[width=0.7\linewidth]{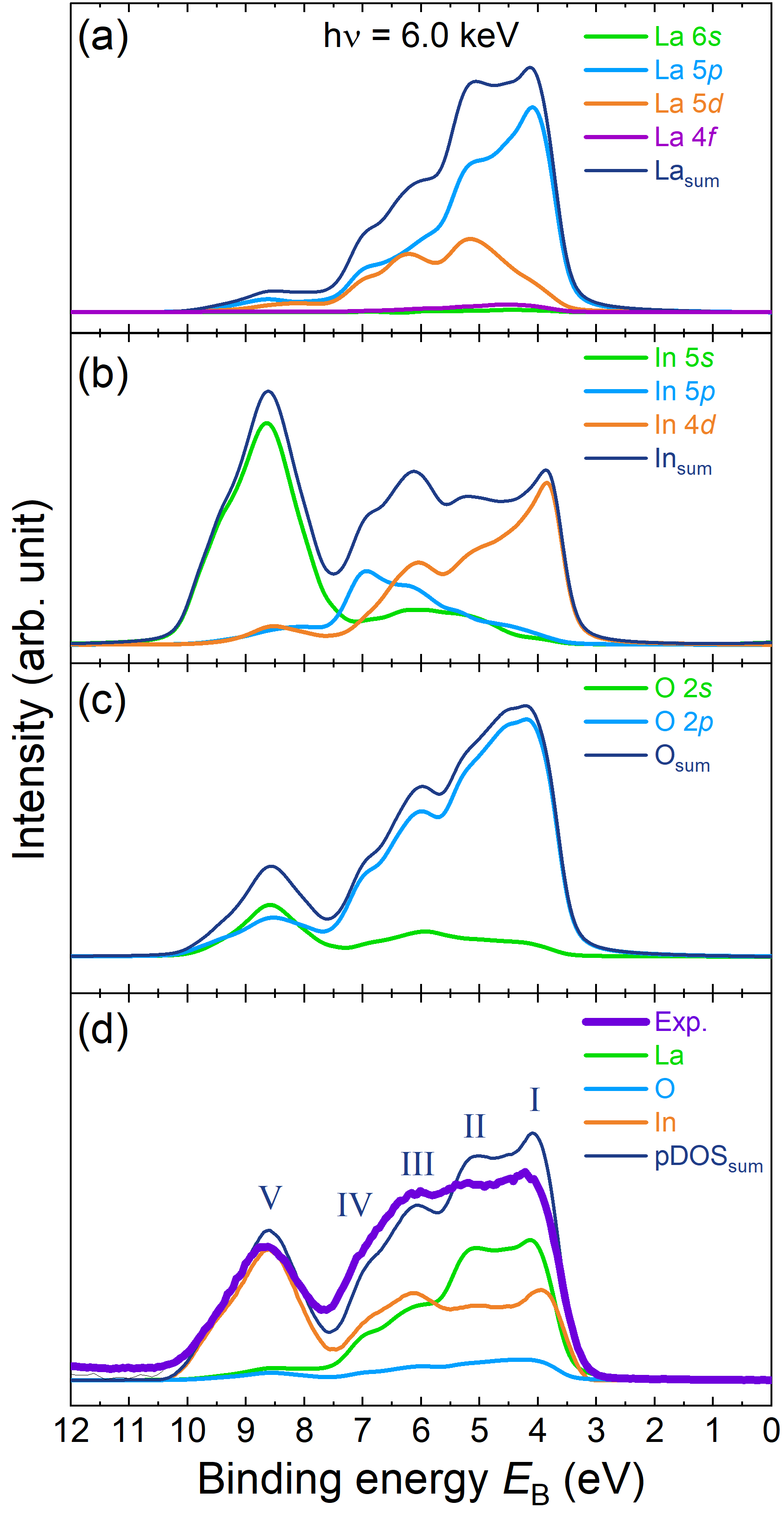}
        \caption{Valence spectra of \ce{LaInO3} showing (a)-(c) element- and orbital-specific one-electron photoionisation cross section corrected projected densities of state (PDOS) from DFT, and (d) total element-specific one-electron photoionisation cross section corrected PDOS and HAXPES data (Exp.) collected at a photon energy of 6.0~keV. In (d), experiment and theory were normalised to the area of the VB spectra and pDOS between 0 and 11~eV, respectively.}
        \label{fig:LIO_VB_6}
    \end{figure}

\subsection{\texorpdfstring{\ce{BaSnO3}}{BaSnO3}/\texorpdfstring{\ce{LaInO3}}{LaInO3} heterostructures}

\subsubsection{Core States}
\begin{figure}
        \centering
        \includegraphics[width=1.0\linewidth]{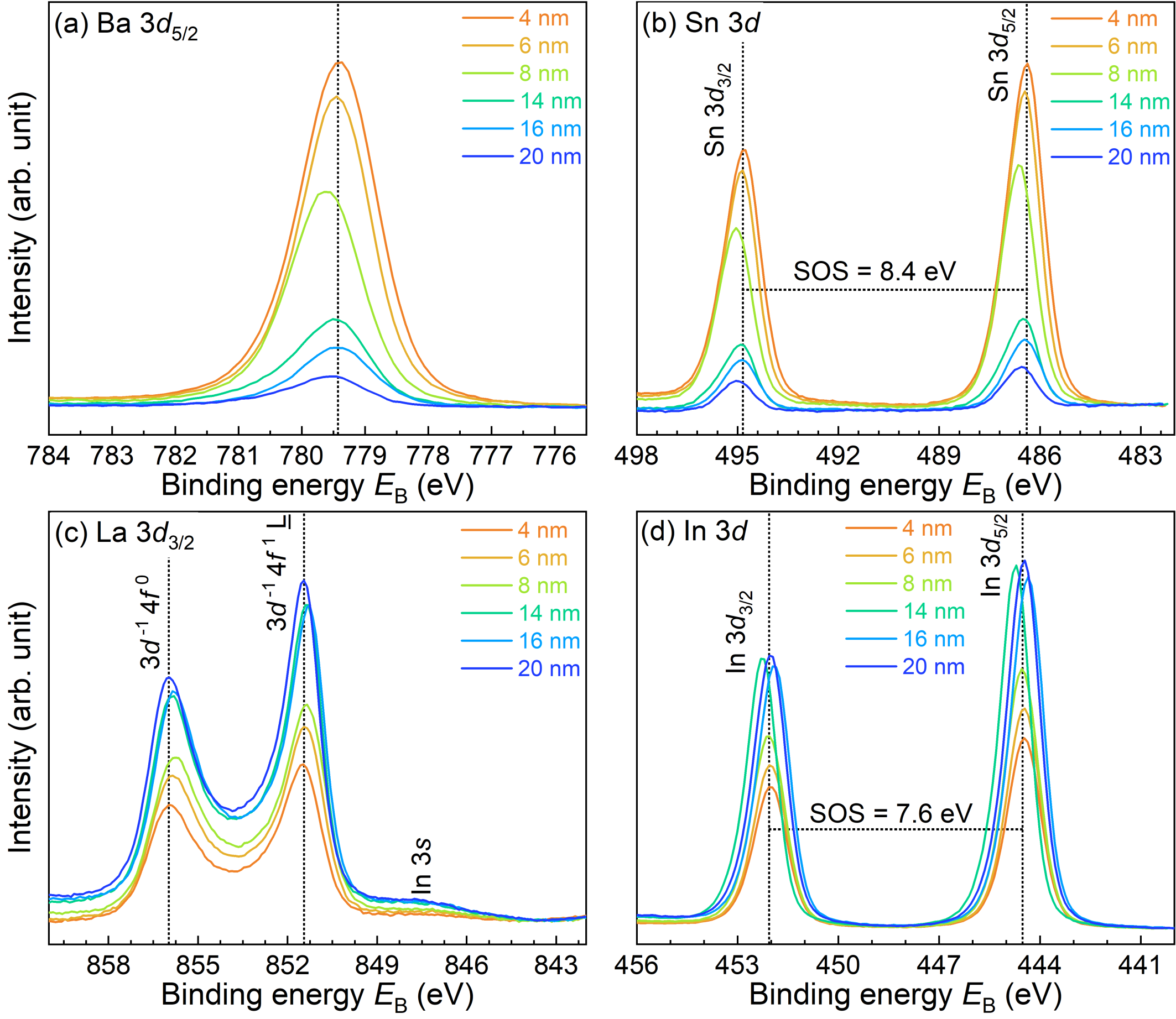}
        \caption{Key core state spectra of \ce{BaSnO3}/\ce{LaInO3} heterostructures collected at a photon energy of 6.0~keV, including (a) Ba~3\textit{d}\textsubscript{5/2}, (b) Sn~3\textit{d}, (c) La~3\textit{d}\textsubscript{3/2}, and (d) In~3\textit{d}. For the Sn and In~3\textit{d} doublets, the value of the experimental spin-orbit splitting (SOS) is included. The spectra are normalised to the sum of the areas of the In~3\textit{d}\textsubscript{5/2} and Sn~3\textit{d}\textsubscript{5/2} lines. Data are aligned to the Au~$E_F$ position for \ce{LaInO3} thicknesses of 14, 16, and 20~nm and to the intrinsic 2DEG feature for \ce{LaInO3} thicknesses of 4, 6, and 8~nm.}
        \label{fig:HS_CLs}
    \end{figure}

The core state spectra of the heterostructures (see Figure~\ref{fig:HS_CLs}) follow the observations for the \ce{BaSnO3} and \ce{LaInO3} references as discussed above. The change in \ce{LaInO3} overlayer thickness leads to obvious systematic changes in the relative peak intensities of core states arising from both layers. In addition, subtle differences in peak positions and line shapes are observable, stemming from variations in the IMFP and effective attenuation length with changes in photon energy (see Tables~S1 and S2 in the Supplementary Information for the estimated IMFPs), thereby altering the probing depth. This affects the relative contributions from the two oxides, as well as their peak positions and line shapes, which change accordingly (see Figures~S8 and S9 in the Supplementary Information for the core state spectra collected at 3.4 and 1.5~keV, respectively). Overall, the close correspondence of the spectral line shapes and $E_B$ to those of the reference samples is consistent with the constituent oxides retaining broadly similar local chemical environments in the heterostructures, although minor interfacial intermixing or segregation below the sensitivity of the measurements cannot be excluded. 

\subsubsection{Semi-core States}
\begin{figure}
        \centering
        \includegraphics[width=1.0\linewidth]{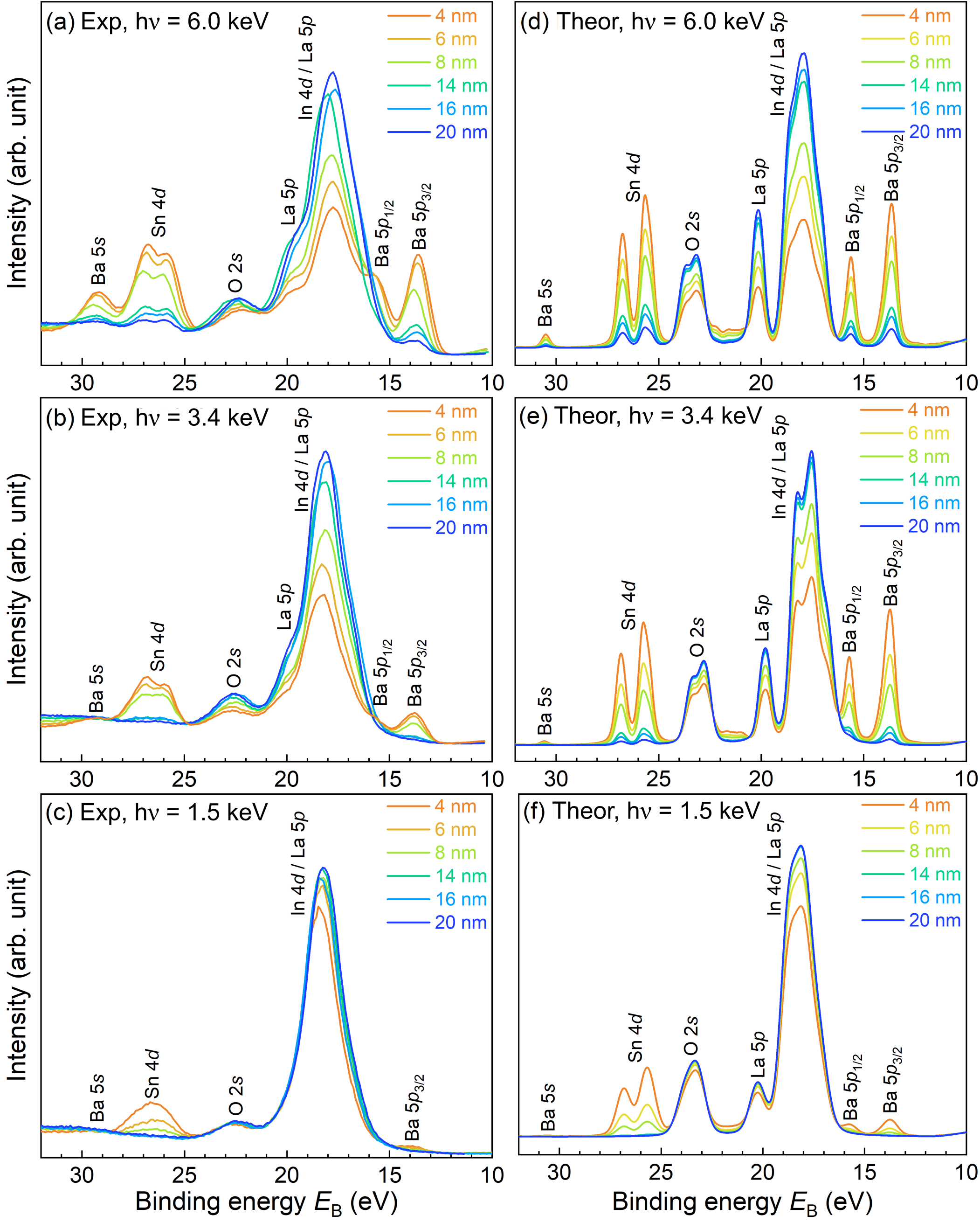}
        \caption{Semi-core state spectra of \ce{BaSnO3}/\ce{LaInO3} heterostructures showing (a)-(c) experimental data (Exp) collected at 6.0, 3.4, and 1.5~keV, respectively, and (d)-(f) densities of states (DOS, Theor) based on element- and orbital-specific one-electron photoionisation cross section corrected projected densities of state (PDOS) from DFT.}
        \label{fig:HS_SCL}
    \end{figure}

The semi-core state spectra of the heterostructures are rather complicated due to the overlap of features from both oxides (see Figure~\ref{fig:HS_SCL}). However, by comparing both experimental data collected at varying photon energies (Figures~\ref{fig:HS_SCL}(a)-(c)) and theoretical PDOS corrected for cross sections at these photon energies (Figures~\ref{fig:HS_SCL}(d)-(f)), as well as building upon the results and discussion for the isolated oxides, the elemental and orbital character of the individual features can be identified. The relative changes in intensity are as expected from \ce{LaInO3} overlayer thickness and photoionisation cross section changes. The PDOS agree well with the experimental data. The Ba~5\textit{s} and 5\textit{p}\textsubscript{3/2} states are well resolved, particularly in the 6.0~keV data due to increasing probing depth and cross sections, and are observed at $E_B$ of 29.2 and 13.6~eV for the 4~nm \ce{LaInO3} overlayer, respectively. The Ba~5\textit{p}\textsubscript{1/2} state overlaps with the more dominant In~4\textit{d} / La~5\textit{p} feature, which has a main peak at 17.8~eV for the same sample and photon energy. The main contribution for Sn, the Sn~4d doublet, appears at 25.9 and 26.8~eV (SOS = 0.9~eV), 3.6~eV below the O~2\textit{s} feature. Whilst all other semi-core features originate from one of the two oxides only, the O~2s is a convolution of contributions from both \ce{BaSnO3} and \ce{LaInO3}. This leads to a subtle narrowing of the total feature width as it becomes more dominated by \ce{LaInO3}.

\subsubsection{Valence States, Band Offset, and Conduction Band Population}

\begin{figure}
        \centering
        \includegraphics[width=1.0\linewidth]{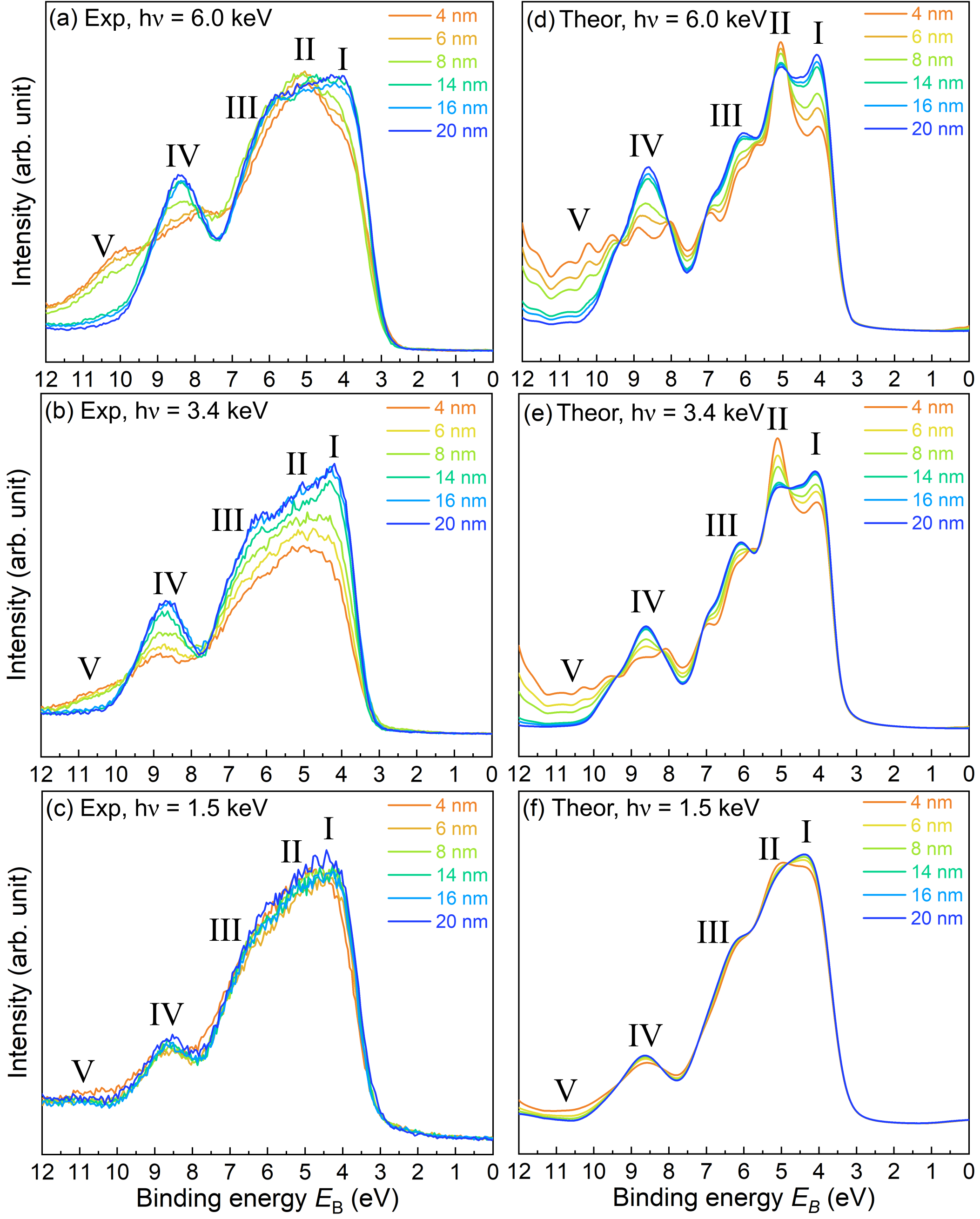}
        \caption{Valence spectra of \ce{BaSnO3}/\ce{LaInO3} heterostructures showing (a)-(c) experimental data (Exp) collected at 6.0, 3.4, and 1.5~keV, respectively, and (d)-(f) densities of states (DOS, Theor) based on element- and orbital-specific one-electron photoionisation cross section corrected projected densities of state (PDOS) from DFT.}
        \label{fig:HS_VB}
    \end{figure}

The valence band (VB) spectra for the heterostructures collected at the three photon energies as well as the corresponding photoionisation cross-section corrected DOS are shown in Figure~\ref{fig:HS_VB}. Experiment and theory agree very well, allowing for discussion of the origin of the observed features. The top of the VB (feature I) is dominated by La~5\textit{p} and In~4\textit{d} states from \ce{LaInO3}. Therefore, as the \ce{LaInO3} thickness decreases the intensity of this feature at 4~eV decreases. Feature III follows the same trend in intensity as it is also dominated by La and In states. In tandem, the relative intensity of feature II at 5~eV, which is dominated by Ba and Sn states, increases with a decrease in \ce{LaInO3} thickness. Finally, a similar intensity modulation with overlayer thickness is observed for the In~5\textit{s} dominated feature IV at $\approx$8.5~eV and the Sn~5\textit{s} dominated feature V at $\approx$10~eV.\\

\begin{figure}
        \centering
        \includegraphics[width=0.5\linewidth]{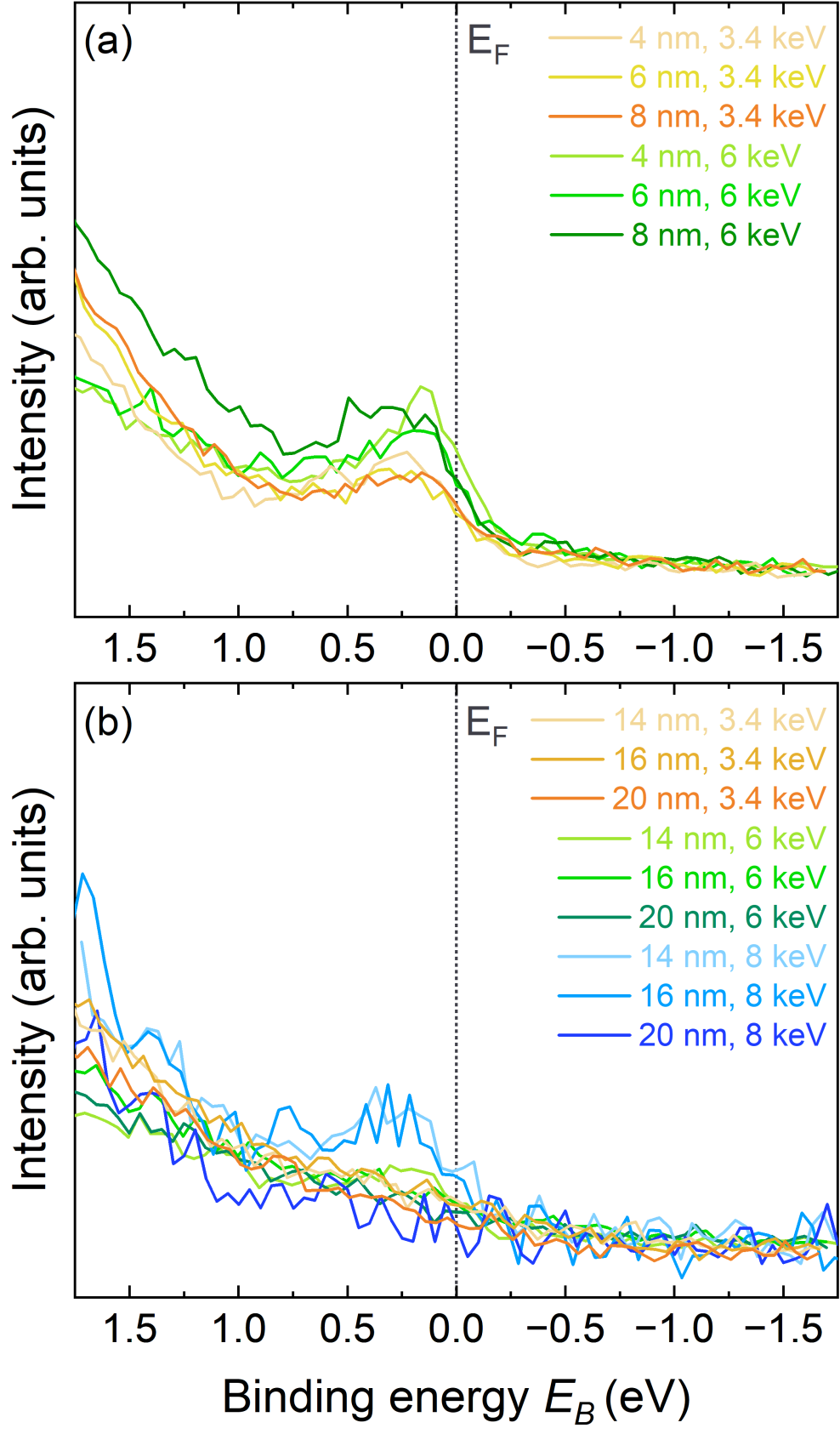}
        \caption{Expanded view of the Sn~5\textit{s} electron population at the Fermi energy $E_F$ of \ce{BaSnO3}/\ce{LaInO3} heterostructures collected at 3.4, 6.0, and 8.0~keV, for the (a) thinner (4 - 8~nm) and (b) thicker (14 - 20~nm) \ce{LaInO3} overlayers, respectively.}
        \label{fig:HS_EF}
    \end{figure}

In the experimental VB spectra presented in Figure~\ref{fig:HS_VB}, no features above the background are observable between the VBM and $E_F$, as might be expected from an occupied 2DEG state. This reflects the very low spectral intensity of the 2DEG-related states relative to the main VB. Figure~\ref{fig:HS_EF} therefore presents a magnified view of the spectra close to $E_F$. Near-$E_F$ intensity attributable to the 2DEG electron population is observed for overlayer thicknesses and photon energies for which the photoelectron attenuation through the \ce{LaInO3} layer is sufficiently low. The observed near-$E_F$ intensity is consistent with occupation of the Sn~5\textit{s}-derived \ce{BaSnO3} conduction band, as predicted theoretically for the interfacial 2DEG.~\cite{Aggoune2021} Its progressive attenuation with increasing \ce{LaInO3} thickness, and its absence above the experimental background for the 20-nm overlayer, demonstrate that the signal originates from a buried region rather than from the \ce{LaInO3} surface or bulk. This depth dependence is consistent with assignment of the occupied states to the \ce{LaInO3}/\ce{BaSnO3} interface, although the photoelectron sectroscopy measurements alone do not precisely determine their spatial extent into the \ce{BaSnO3} layer. The interfacial assignment is further supported by complementary experimental and theoretical results previously reported for the same and related systems.~\cite{Hoffmann2024,Berens2020}

\begin{figure}
        \centering
        \includegraphics[width=0.65\linewidth]{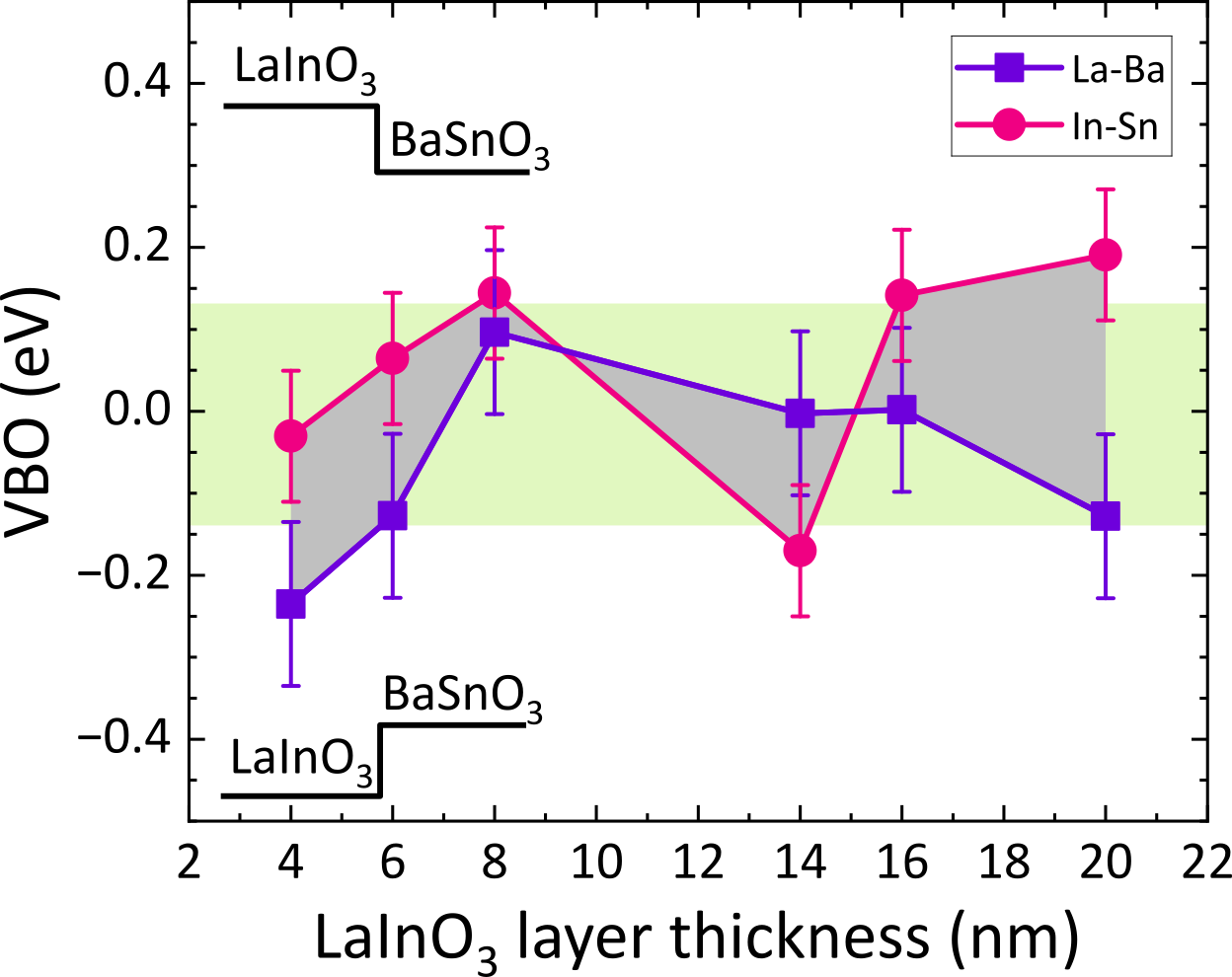}
        \caption{Valence band offset (VBO) values extracted from the experimental spectral positions at 6~keV for the La-Ba and In-Sn pairs for the \ce{BaSnO3}/\ce{LaInO3} heterostructures with varying \ce{LaInO3} thickness, \textit{d}. The green shaded area indicates the range of average VBO values determined from the two core state pairs used for analysis. The following equations illustrate the combination of spectral data used to determine the presented values; La-Ba:
     $(\mathrm{LaInO}_3\mathrm{}_{\mathrm{La}}^{\mathrm{ref}}-\mathrm{LaInO}_3\mathrm{}_{\mathrm{VB}}^{\mathrm{ref}})
-(\mathrm{BaSnO}_3\mathrm{}_{\mathrm{Ba}}^{\mathrm{ref}}-\mathrm{BaSnO}_3\mathrm{}_{\mathrm{VB}}^{\mathrm{ref}})
-(\mathrm{BaSnO}_3/\mathrm{LaInO}_3\mathrm{}_{\mathrm{La}}-\mathrm{BaSnO}_3/\mathrm{LaInO}_3\mathrm{}_{\mathrm{Ba}})$ and In-Sn: $(\mathrm{LaInO}_3\mathrm{}_{\mathrm{In}}^{\mathrm{ref}}-\mathrm{LaInO}_3\mathrm{}_{\mathrm{VB}}^{\mathrm{ref}})
-(\mathrm{BaSnO}_3\mathrm{}_{\mathrm{Sn}}^{\mathrm{ref}}-\mathrm{BaSnO}_3\mathrm{}_{\mathrm{VB}}^{\mathrm{ref}})
-(\mathrm{BaSnO}_3/\mathrm{LaInO}_3\mathrm{}_{\mathrm{In}}-\mathrm{BaSnO}_3/\mathrm{LaInO}_3\mathrm{}_{\mathrm{Sn}})$. The sketches in the upper and lower left corner are guides to the eyes for positive and negative VBO values with respect to the relative \ce{BaSnO3} and \ce{LaInO3} VBM.
        }
        \label{fig:Kraut}
    \end{figure}

Using the Kraut method the valence band offset (VBO) at the \ce{BaSnO3}/\ce{LaInO3} interface could be determined.~\cite{Kraut1980} This method uses the core level $E_B$ and valence band maxima (VBM) for each bulk material and in combination with differences in the core level $E_B$ of the heterostructures allows the extraction of the VBO. Figure~\ref{fig:Kraut} provides an overview of the so extracted values based on the La-Ba and In-Sn core level pairs. Following the definition of VBO = VBM(\ce{LaInO3})-VBM(\ce{BaSnO3}), in the literature, VBO values of $\approx$~0.1~eV calculated from theory,~\cite{Krishnaswamy2016,Aggoune2021} and VBO~=~-0.3~eV derived from experiment are reported.~\cite{Kim2015} In the latter case, optical absorption experiments and electrical characterisation, i.e.\ Fowler-Nordheim tunnelling processes were combined for extraction of the band gaps and conduction band offset, respectively. Based on the average of the VBOs determined from the two possible combinations of core state spectra at comparable probing depth, i.e. Ba~3\textit{d} – La~3\textit{d} and Sn~3\textit{d} – In~3\textit{d}, the present results confirm a VBO range of –0.14 to +0.13~eV for the heterostructures with varying \ce{LaInO3} overlayer thickness, in line with literature reports.

Although these experimental results are consistent with the theoretical predictions, two points should be noted. (i) In Ref.~\cite{Aggoune2021}, the VBO is derived from a comparison of O~\textit{2p} states, whereas in the present work it is determined from metal~\textit{d} states. The resulting values may differ slightly due to differences in screening effects between the O~\textit{2p} and metal~\textit{d} states. (ii) The band offsets reported in Ref.~\cite{Aggoune2021} are based on an idealised system exhibiting a built-in potential extending from the interface to the surface. In the experimentally grown \ce{BaSnO3}/\ce{LaInO3} heterostructures, compensation of this built-in potential at the surface is a plausible scenario and would, in turn, affect the band alignment.

\par

\section{Conclusions}

This work presents a comprehensive investigation of the electronic structure of \ce{BaSnO3}, \ce{LaInO3}, and \ce{BaSnO3}/\ce{LaInO3} heterostructures with systematically varied \ce{LaInO3} overlayer thicknesses by combining soft and hard X-ray photoelectron spectroscopy with hybrid density functional theory calculations. The excellent agreement between experiment and photoionisation-cross-section-corrected projected densities of states enables a detailed assignment of core, semi-core, and valence band features and provides a consistent picture of the chemical bonding and orbital hybridisation in both parent oxides.

For the heterostructures, the evolution of spectral weight with \ce{LaInO3} thickness and photon energy is fully consistent with the designed layer architecture and probing depths, allowing the respective contributions from \ce{BaSnO3} and \ce{LaInO3} to be disentangled. Importantly, hard X-ray photoelectron spectroscopy directly reveals the presence of occupied electronic states at the Fermi level associated with the interfacial two-dimensional electron gas, and their photon energy dependent visibility is consistent with these states being spatially confined to the \ce{BaSnO3}/\ce{LaInO3} interface. Using the Kraut method, the valence band offset could be estimated as –0.15 to +0.12~eV across the heterostructure series. This direct spectroscopic observation complements previous electrical and structural studies and provides compelling evidence for interface-driven electronic reconstruction in this system.

Overall, this study establishes a detailed electronic structure reference for \ce{BaSnO3}/\ce{LaInO3} heterostructures and demonstrates the power of combining depth-resolved photoelectron spectroscopy with advanced first-principles calculations. The insights gained here form a robust foundation for future investigations of interface engineering, carrier confinement, and functional optimisation in high-mobility oxide heterostructures.

\begin{acknowledgments}
AAR and CK acknowledge the support from the Department of Chemistry, UCL. This work was partially performed in the framework of GraFOx, a Leibniz ScienceCampus partially funded by the Leibniz Association. GH gratefully acknowledges ﬁnancial support from the Leibniz-Gemeinschaft under Grant No. K74/2017. The authors acknowledge DESY (Hamburg, Germany), a member of the Helmholtz Association HGF, for the provision of experimental facilities. Parts of this research were carried out at PETRA III using beamline P22 for proposal I-20221263. Funding for the HAXPES instrument by the Federal Ministry of Education and Research (BMBF) under framework program ErUM is gratefully acknowledged. WA acknowledges funding from the Deutsche Forschungsgemeinschaft (DFG, German Research Foundation) under project No. 510995508. WA gratefully acknowledges the computing time granted by the Resource Allocation Board and provided on the supercomputer Lise and Emmy at NHR@ZIB and NHR@Göttingen as part of the NHR infrastructure.

\end{acknowledgments}

\section*{Disclosures}
The authors declare no conflicts of interest.

\section*{Data Availability}
The DFT input and output files can be downloaded free of charge from the NOMAD Repository~\cite{draxl-19jpm} at the following link: \url{https://dx.doi.org/10.17172/nomad.0w1z-nz4m}. 

\bibliography{references.bib}
\bibliographystyle{apsrev4-1}

\end{document}


\preprint{APS/123-QED}

\title[PRB]{Electronic structure, band offset, and interface electron population of the \texorpdfstring{\ce{LaInO3}}{LaInO3}/\texorpdfstring{\ce{BaSnO3}}{BaSnO3} system}

\author{G.~Hoffmann}
\affiliation{Paul-Drude-Institut für Festkörperelektronik, Leibniz-Institut im Forschungsverbund Berlin e.\,V., Hausvogteiplatz 5--7, 10117 Berlin, Germany.}

\author{A.~A.~Riaz}
\affiliation{Department of Chemistry, University College London, 20 Gordon Street, London, WC1H~0AJ, United Kingdom.}

\author{C.~Kalha}
\affiliation{Department of Chemistry, University College London, 20 Gordon Street, London, WC1H~0AJ, United Kingdom.}

\author{A.~Gloskovskii}
\author{C.~Schlueter}
\affiliation{Deutsches Elektronen-Synchrotron DESY, Notkestrasse 85, 22607 Hamburg, Germany.}

\author{O.~Brandt}
\affiliation{Paul-Drude-Institut für Festkörperelektronik, Leibniz-Institut im Forschungsverbund Berlin e.\,V., Hausvogteiplatz 5--7, 10117 Berlin, Germany.}

\author{W.~Aggoune}
\author{C.~Draxl}
\affiliation{Institut f{\"u}r Physik and CSMB, Humboldt-Universit{\"a}t zu Berlin, 12489 Berlin, Germany.}

\author{O.~Bierwagen}
\affiliation{Paul-Drude-Institut für Festkörperelektronik, Leibniz-Institut im Forschungsverbund Berlin e.\,V., Hausvogteiplatz 5--7, 10117 Berlin, Germany.}

\author{A.~Regoutz}
 \email{anna.regoutz@chem.ox.ac.uk}
 \affiliation{Department of Chemistry, University of Oxford, Inorganic Chemistry Laboratory, South Parks Road, Oxford, OX1~3QR, United Kingdom.}
\affiliation{Department of Chemistry, University College London, 20 Gordon Street, London, WC1H~0AJ, United Kingdom.}

\date{\today}
\maketitle

\newpage

 \tableofcontents

\cleardoublepage

\section{Electrical Characterisation}

\begin{figure*}[h]
    \centering
    \includegraphics[width=1\linewidth]{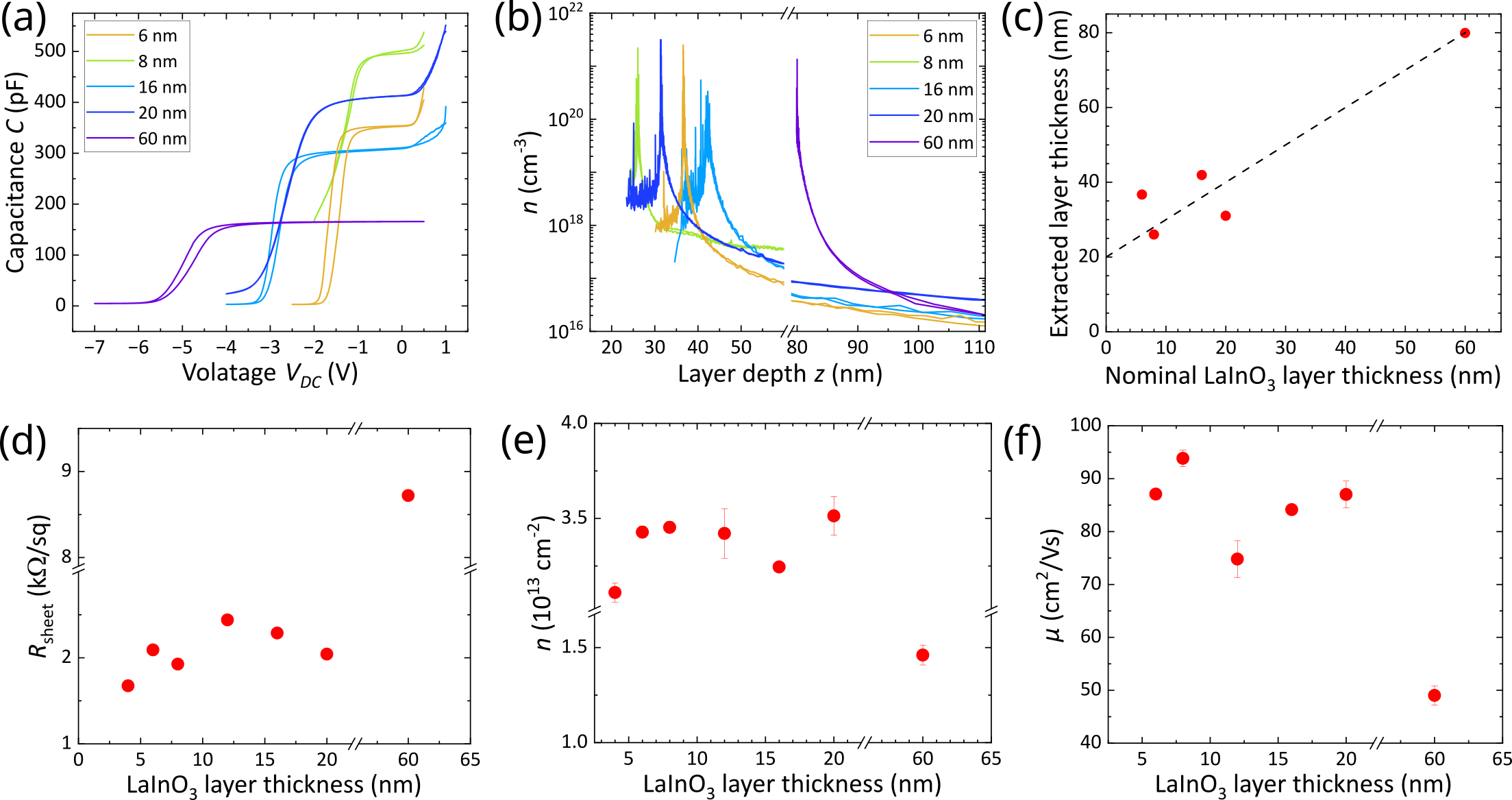}
    \caption{Electrical properties derived from capacitance voltage (CV), van der Pauw (vdP) and Hall measurements. (a) Representative CV curves of some samples investigated in this study. (b) Carrier concentration profiles as a function of layer thickness extracted from CV curves shown in (a). Details on experimental setup and measurements can be extracted from Ref.~\cite{Hoffmann2024}. (c) Nominal vs. extracted LaInO$_3$ layer thickness from profiles in (b). The black dashed line indicates the expected relation with respect to an offset of 20~nm. (d)-(f) Sheet resistance, carrier concentration, and mobility extracted from vdP and Hall measurements on these samples, respectively.\\}
    \label{fig:electrical_properties}
\end{figure*}

Fig.~\ref{fig:electrical_properties}(a) shows representative capacitance-voltage (CV) curves of some of the samples investigated in this study. The plateau at negative applied voltages indicates the presence of a 2DEG. In Fig.~\ref{fig:electrical_properties}(b) the profiles of (a) were translated into carrier concentration profiles as a function of film thickness. For all samples, a carrier accumulation of a few nm width can be observed which we attribute to the presence of a 2DEG at the interface. Fig.~\ref{fig:electrical_properties}(c) compares the nominal peak positions to the observed ones in (b). A black dashed line serves as guide to the eyes regarding the expected values modulo an offset of 20~nm. We attribute the offset as well as deviation from the dashed line to an increased layer thickness caused by an oxide layer on the Hg surface with fluctuating thickness as well as variations of the Hg contact size that depends on the applied vacuum, respectively. In Fig.~\ref{fig:electrical_properties}(d)-(f), similar sheet resistance, carrier concentration and mobility of the samples with LIO layer thickness between 4 and 20~nm underline the reproducibility of the samples. The LIO reference sample has a higher sheet resistance as well as a lower carrier concentration and mobility. The reduce carrier concentration for thick LaInO$_3$ films has also been observed by others and where described by a comprehensive model (see Kim, Y.M., Markurt, T., Kim, Y. et al. Interface polarization model for a 2-dimensional electron gas at the \ce{BaSnO3}/\ce{LaInO3} interface. Sci Rep 9, 16202 (2019)).

\cleardoublepage

\section{AFM and XRD}

\begin{figure*}[h]
    \centering
    \includegraphics[width=1\linewidth]{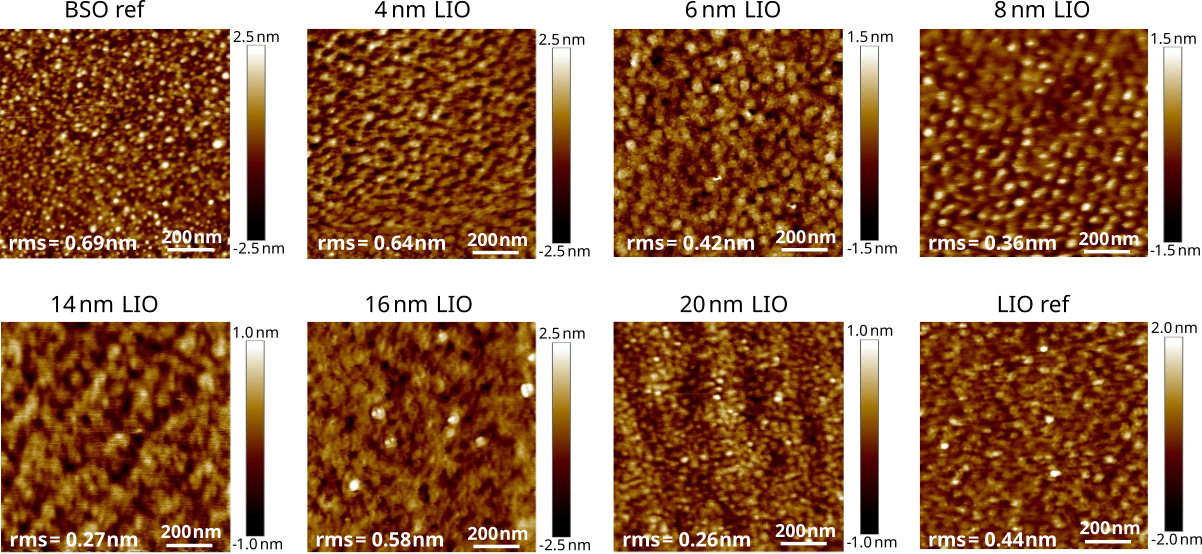}
    \caption{AFM images of the sample surface of a 1$\times$1~$\mu$µm area for all samples investigated in this study. The roughness is given by the root mean square (rms) value. }
    \label{fig:AFM_images}
\end{figure*}

\begin{figure}[h]
    \centering
    \includegraphics[width=0.5\linewidth]{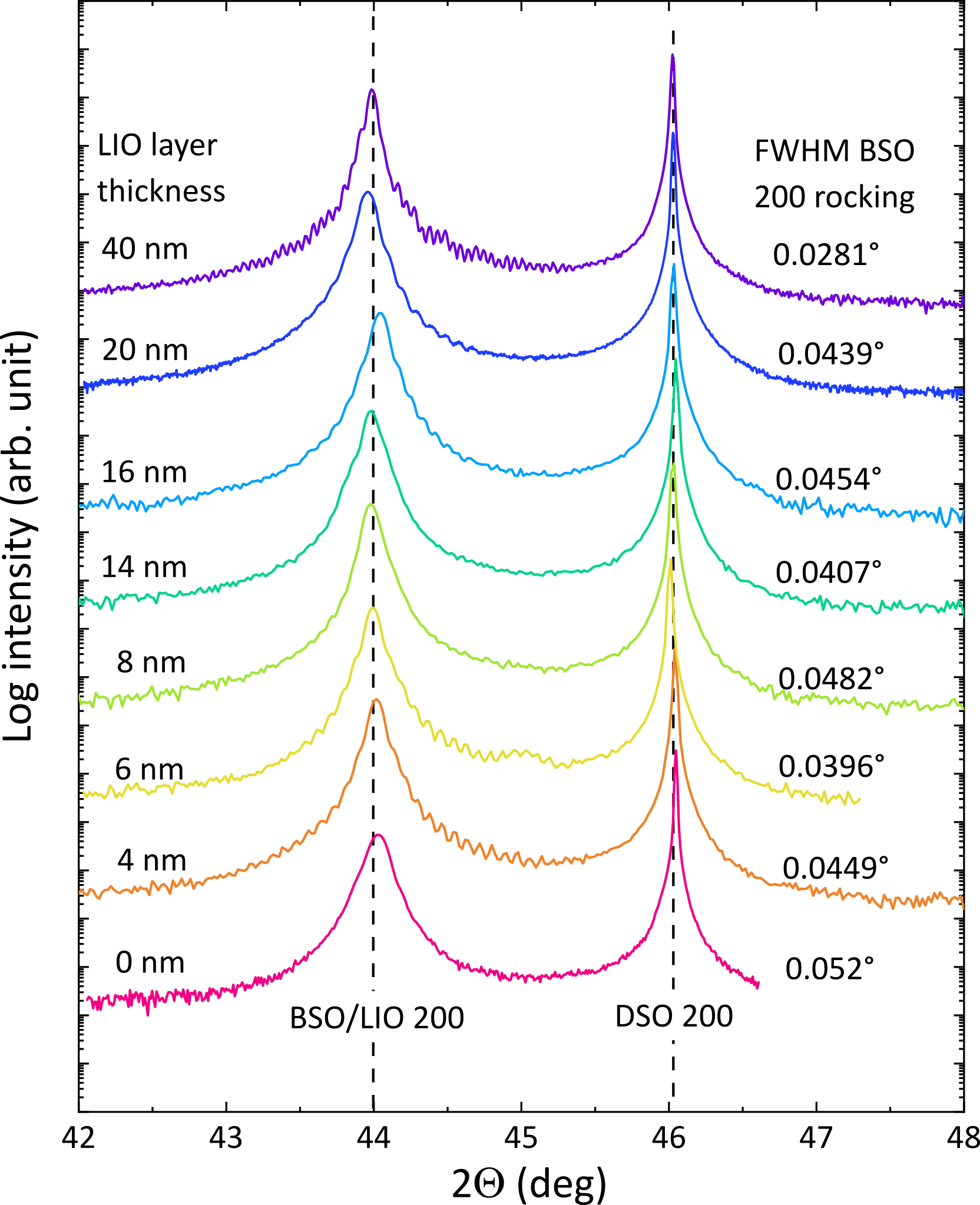}
    \caption{XRD 2$\Theta - \omega$ scan of the BSO/LIO 200 peak and the DSO 200 peak for all samples. The LIO layer thickness is written on top of the XRD measurements. The data was shifted by a constant offset for better visualisation. The additional peak around 45$^{\circ}$ originates from slightly Sn-rich growth conditions. Full width at half maximum (FWHM) values of the BSO 200 rocking curve are written on the right side on top of the corresponding XRD data. Due to the lattice match of BSO and LIO, the films cannot be distinguished in the XRD scans shwon here. Further, the FWHM value of the BSO/LIO 200 reflex is dominated by the BSO layer due to the thickness of $\approx.$~100~nm compared to the thin LIO layers, and the rotational domains that have been observed in LIO layers causing a higher FWHM value around 0.1$^{\circ}$. Note that for the LIO reference sample, the period of the thickness fringes is increased due to the fact that both layers of the same lattice constant are contributing to an overall thickness.}
    \label{fig:xrd_overview}
\end{figure}

\cleardoublepage

\section{Survey Spectra}

\begin{figure*}[ht!]
\centering
    \includegraphics[keepaspectratio, width = 0.7\linewidth]{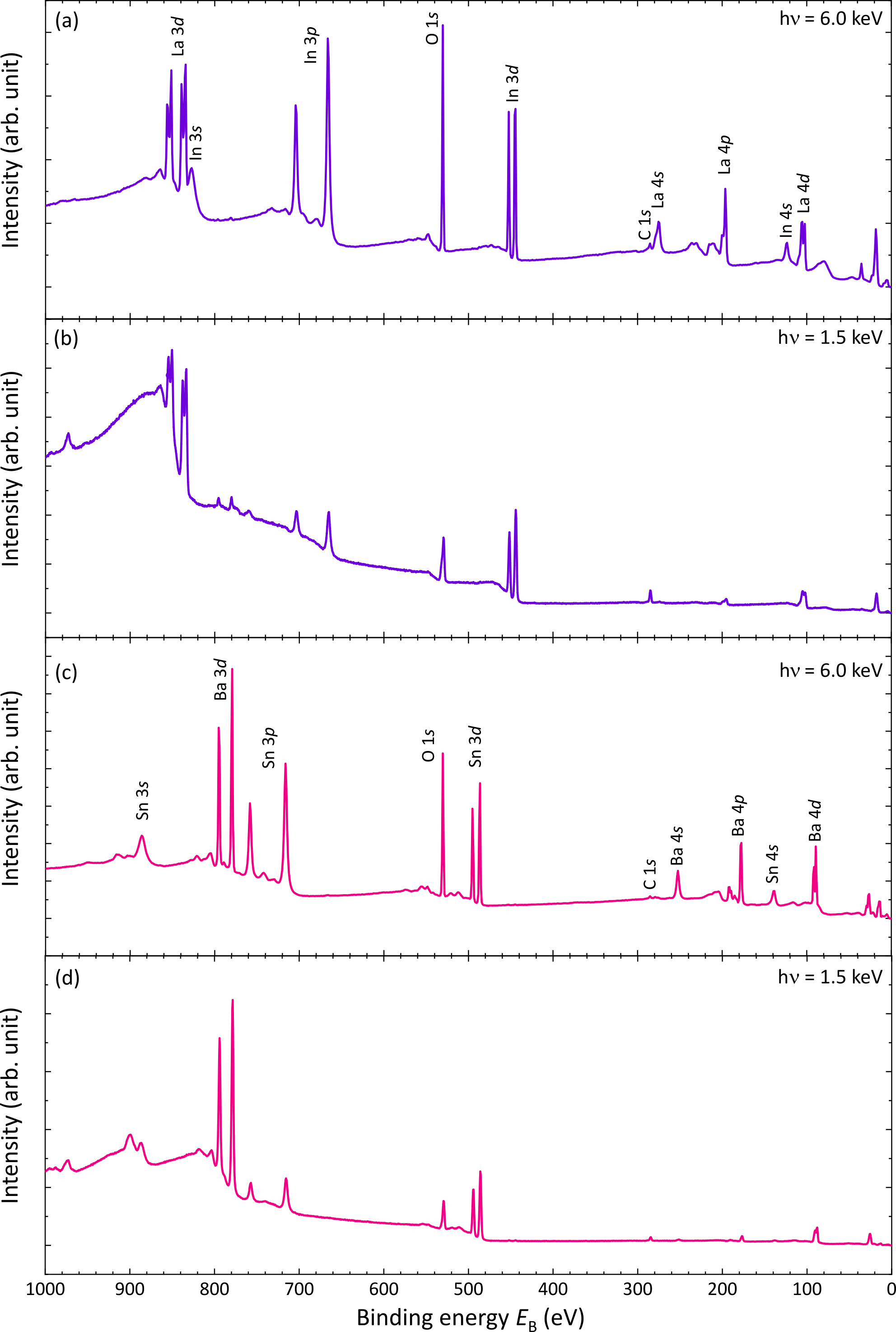}
    \caption{Survey spectra for the (a) and (b) \ce{LaInO3} and (c) and (d) \ce{BaSnO3} reference samples collected at photon energies of (a) and (c) 6.0~keV, and (b) and (d) 1.5~keV. All major core lines are labelled in (a) and (c).}
    \label{fig:HS_survey}
\end{figure*}

\begin{figure*}[ht!]
\centering
    \includegraphics[keepaspectratio, width = 0.7\linewidth]{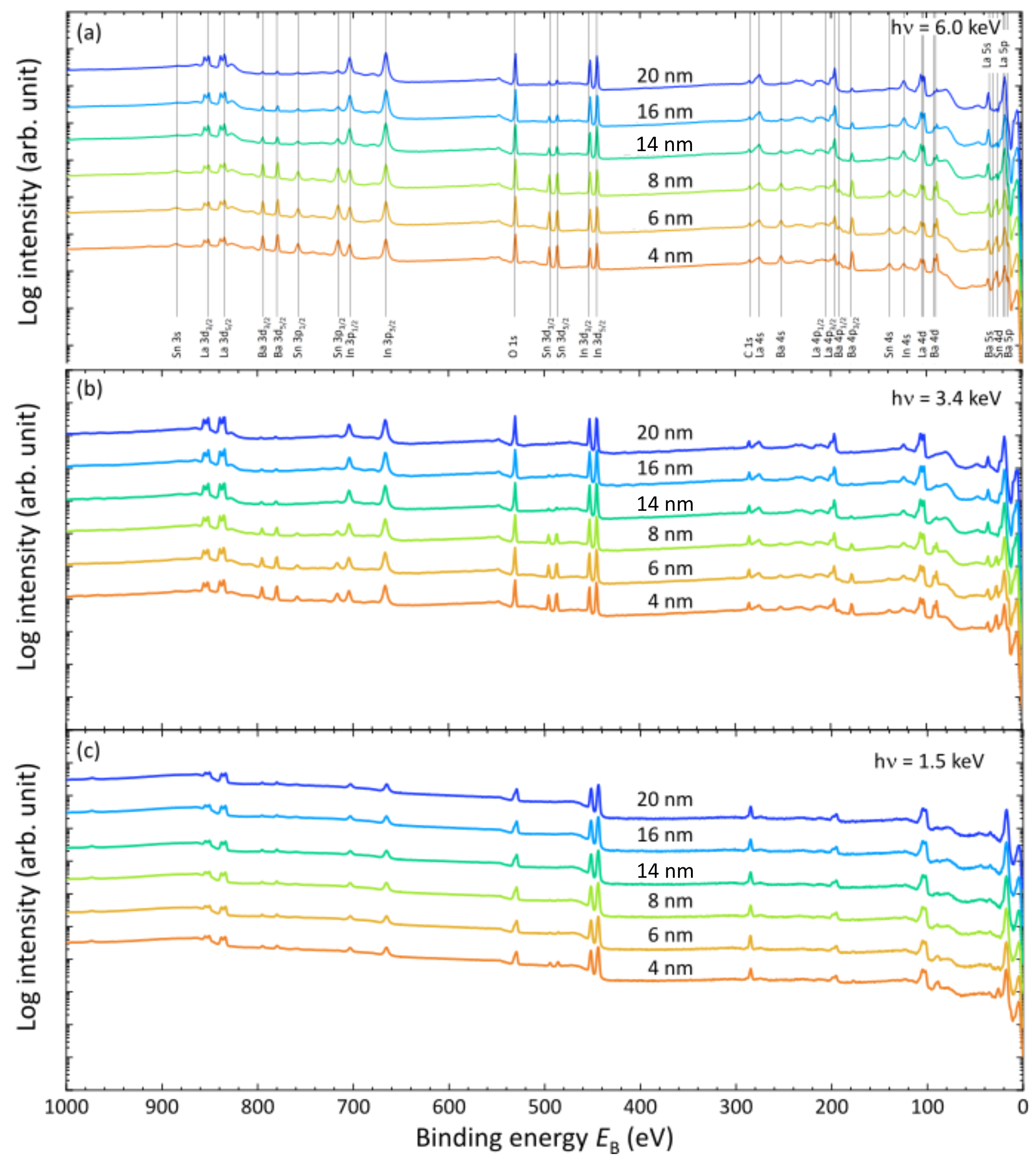}
    \caption{Survey spectra for the \ce{BaSnO3}/\ce{LaInO3} heterostructures with \ce{LaInO3} overlayer thicknesses from 4 - 20~nm collected at photon energies of (a) 6.0~keV, (b) 3.4~keV, and (c) 1.5~keV. All major core lines are labelled in (a) with the same labelling applicable to all photon energies. Spectra are stacked and plotted on a logarithmic intensity scale for ease of viewing.}
    \label{fig:HS_survey2}
\end{figure*}

\cleardoublepage

\section{Inelastic Mean Free Path}

\begin{table}[ht!]
    \centering
    \caption{Material parameters used to calculate the relativistic inelastic mean free path (IMFP) for \ce{LaInO3} and \ce{BaSnO3} using the TPP-2M model as implemented in the QUASES software package,~\cite{Shinotsuka2015} including the bulk density $\rho$,\cite{Villars2023,Villars2023_2} the number of valence electrons $N_v$ as derived in Ref.~\cite{Tanuma2003}, the atomic weight M, and the band gap $E_g$.}
    \begin{tabular}{lccccc}
        \hline\hline

        & $\rho$ / g sm$^{-3}$ & $N_v$, \textit{s}+\textit{p}(lbe) & $N_v$, \textit{d} & M & $E_g$ / eV \\ \hline
         \ce{LaInO3} & 7.19 & 23 & 1 & 301.72047 & 4.3 \\
         \ce{BaSnO3} & 7.33 & 24 & 0 & 304.034 & 3.0 \\
        \hline \hline
    \end{tabular}
    \label{Tab:IMFP_param}
\end{table}

\begin{table}[ht!]
    \centering
    \caption{Maximum relativistic inelastic mean free paths (IMFP) for \ce{LaInO3} and \ce{BaSnO3} at kinetic energies equivalent to the photon energies ($h\nu$) calculated using the TPP-2M model as implemented in the QUASES software package.~\cite{Shinotsuka2015} The material parameters given in Table~\ref{Tab:IMFP_param}.}
    \begin{tabular}{lcc}
        \hline\hline
        & h$\nu$ / keV & IMFP / \AA \\ \hline
         \ce{LaInO3} & 1.5 & 25.6 \\
                     & 3.4 & 48.9 \\
                     & 6.0 & 77.6 \\
                     & 8.0 & 98.2 \\
         \ce{BaSnO3} & 1.5 & 25.1 \\
                     & 3.4 & 47.8 \\
                     & 6.0 & 75.8 \\
                     & 8.0 & 96.0 \\
        \hline \hline
    \end{tabular}
    \label{Tab:IMFP}
\end{table}

\cleardoublepage

\section{Additional Core State Spectra}

\begin{figure*}[ht!]
\centering
    \includegraphics[keepaspectratio, width = 0.35\linewidth]{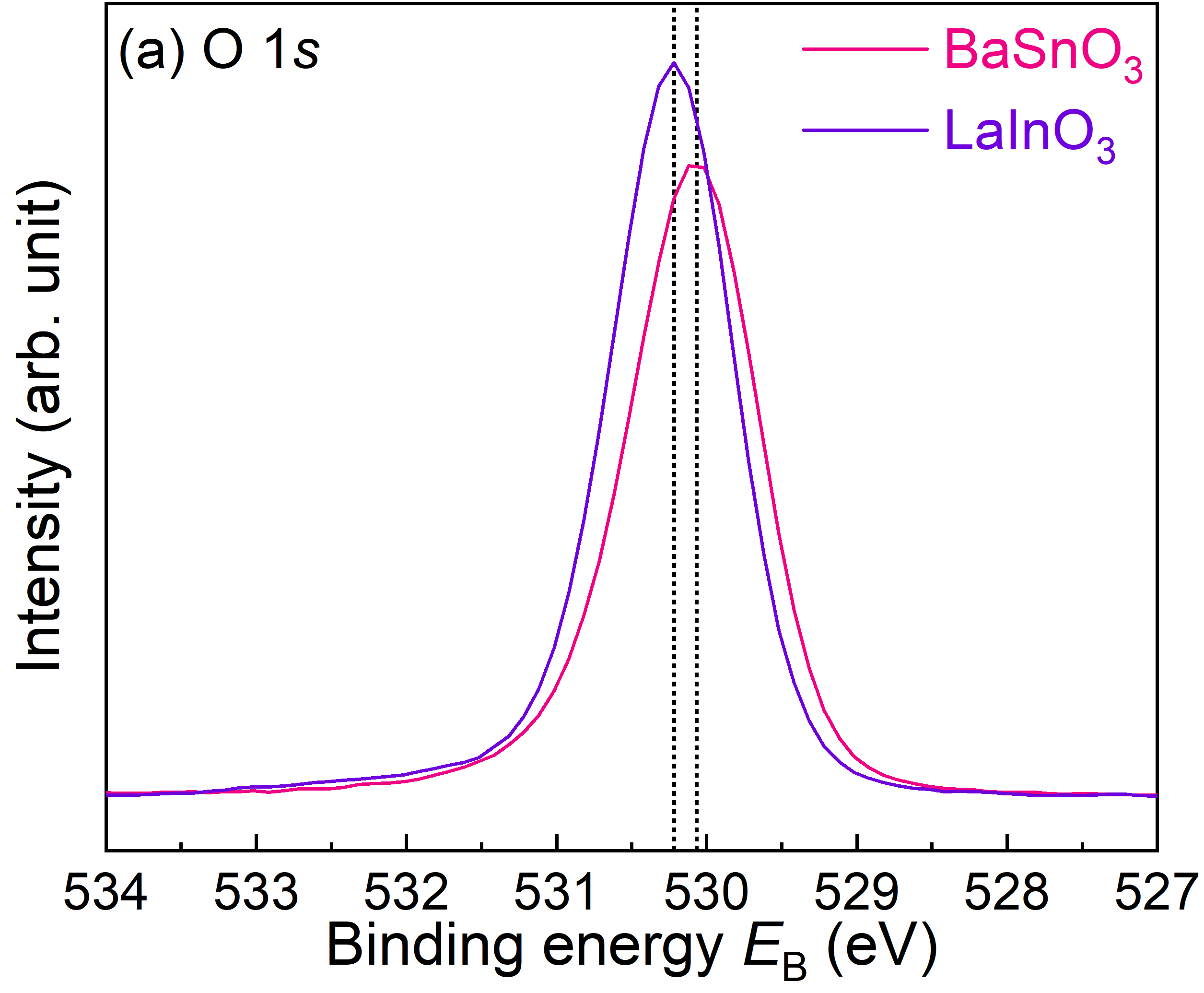}
    \caption{O~1\textit{s} spectra for the \ce{BaSnO3} and \ce{LaInO3} references collected at a photon energy of 6.0~keV.}
    \label{fig:BSO_LIO_6keV_O1s}
\end{figure*}

\begin{figure*}[ht!]
\centering
    \includegraphics[keepaspectratio, width = 0.6\linewidth]{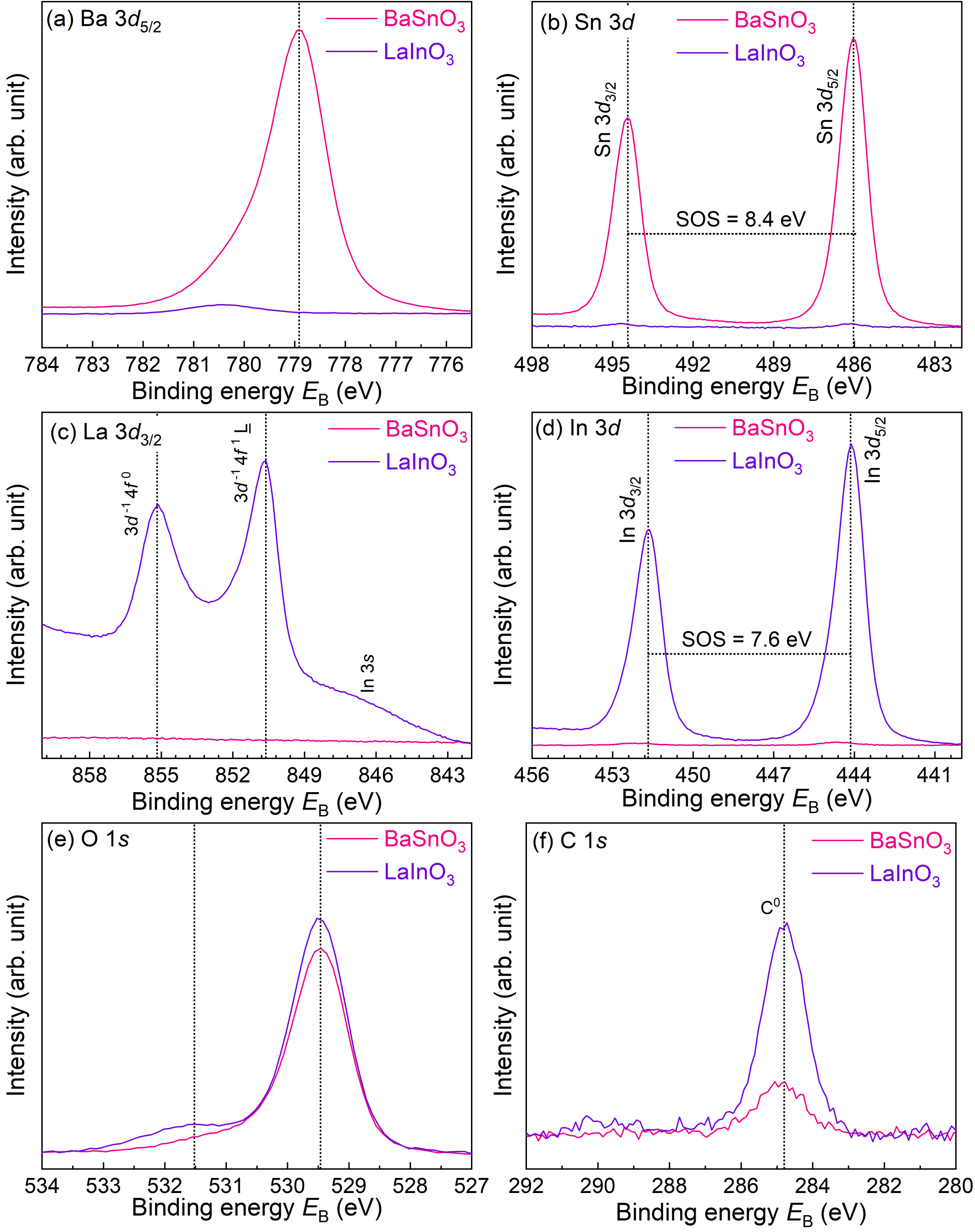}
    \caption{Key core state spectra of \ce{BaSnO3} and \ce{LaInO3} collected at a photon energy of 1.5~keV, including (a) Ba~3\textit{d}\textsubscript{5/2}, (b) Sn~3\textit{d}, (c) La~3\textit{d}\textsubscript{3/2}, (d) In~3\textit{d}, (e) O~1\textit{s}, and (f) C~1\textit{s}. For the Sn and In~3\textit{d} doublets, the spin-orbit splitting (SOS) is included.}
    \label{fig:BSO_LIO_SX_CLs}
\end{figure*}

    \begin{figure}[ht!]
        \centering
        \includegraphics[width=0.6\linewidth]{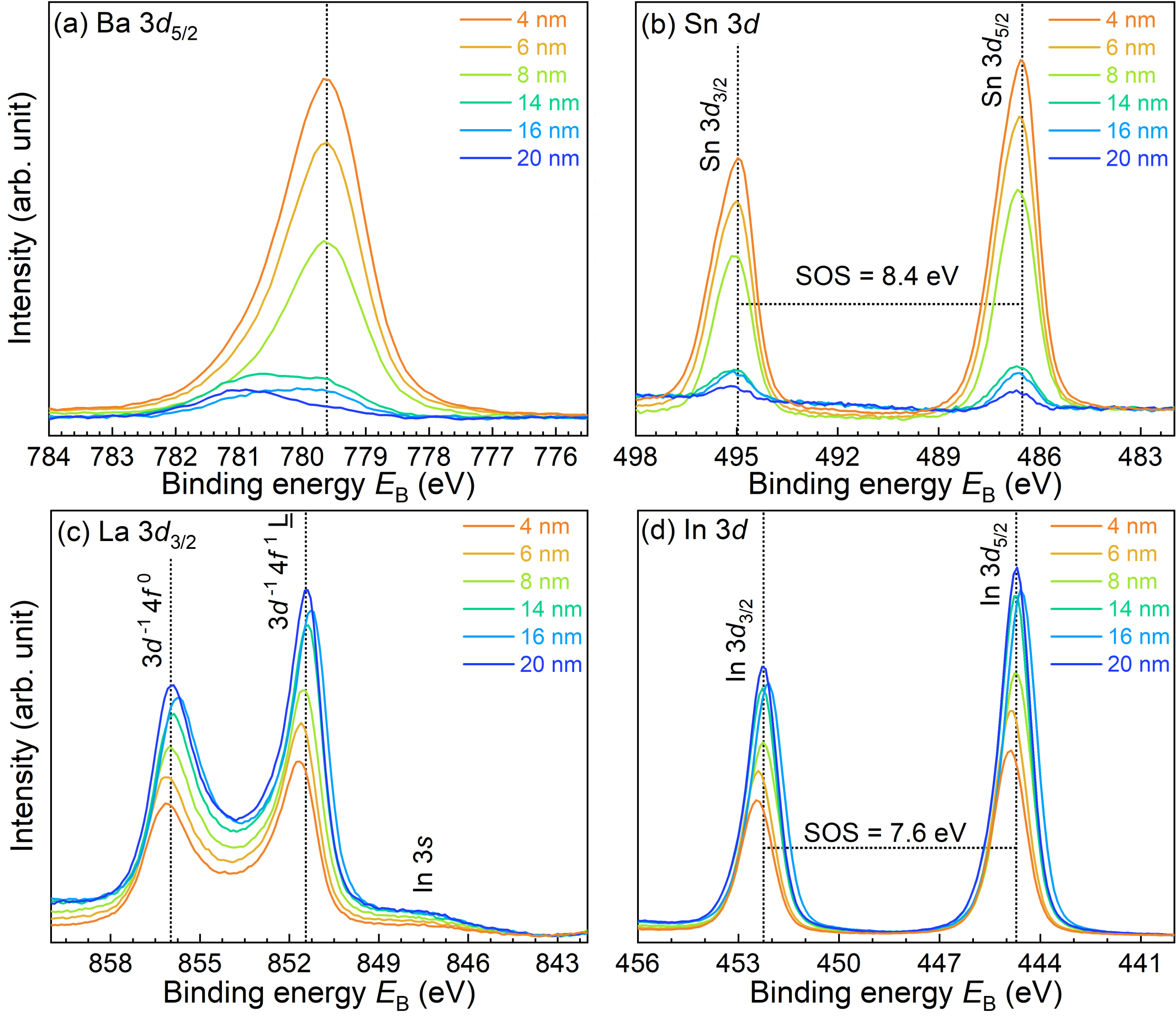}
        \caption{Key core state spectra of \ce{BaSnO3}/\ce{LaInO3} heterostructures collected at a photon energy of 3.4~keV, including (a) Ba~3\textit{d}\textsubscript{5/2}, (b) Sn~3\textit{d}, (c) La~3\textit{d}\textsubscript{3/2}, and (d) In~3\textit{d}. For the Sn and In~3\textit{d} doublets, the spin-orbit splitting (SOS) is included. The spectra are normalised to the sum of the areas of the In~3\textit{d}\textsubscript{5/2} and Sn~3\textit{d}\textsubscript{5/2} lines. Data are aligned to the Au~$E_F$ position for \ce{LaInO3} thicknesses of 14, 16, and 20~nm and to the intrinsic 2DEG feature for \ce{LaInO3} thicknesses of 4, 6, and 8~nm.}
        \label{fig:HS_CLs_3keV}
    \end{figure}

        \begin{figure}[ht!]
        \centering
        \includegraphics[width=0.6\linewidth]{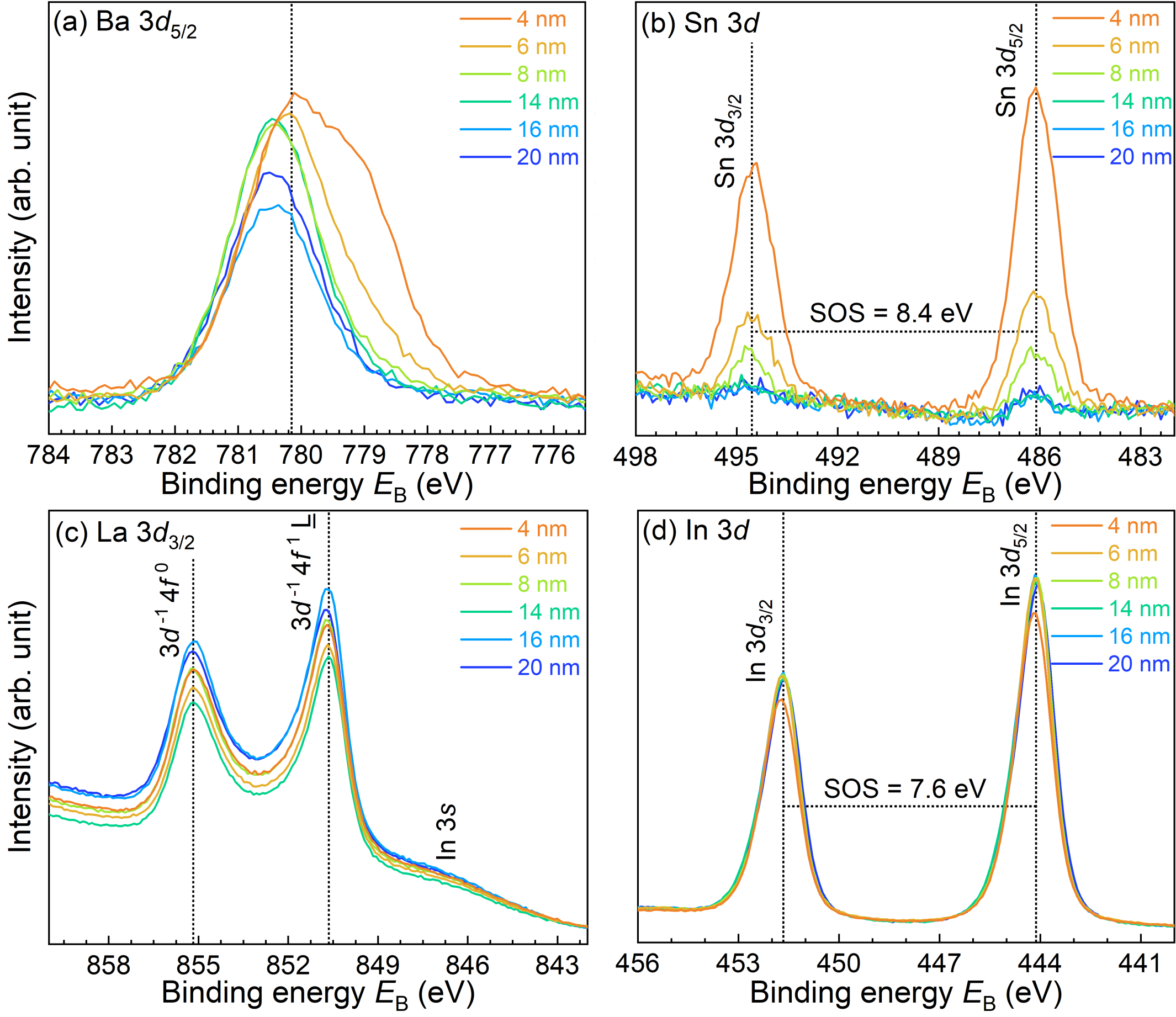}
        \caption{Key core state spectra of \ce{BaSnO3}/\ce{LaInO3} heterostructures collected at a photon energy of 1.5~keV, including (a) Ba~3\textit{d}\textsubscript{5/2}, (b) Sn~3\textit{d}, (c) La~3\textit{d}\textsubscript{3/2}, and (d) In~3\textit{d}. For the Sn and In~3\textit{d} doublets, the spin-orbit splitting (SOS) is included. The spectra are normalised to the sum of the areas of the In~3\textit{d}\textsubscript{5/2} and Sn~3\textit{d}\textsubscript{5/2} lines. Data are aligned to the Au~$E_F$ position for \ce{LaInO3} thicknesses of 14, 16, and 20~nm and to the intrinsic 2DEG feature for \ce{LaInO3} thicknesses of 4, 6, and 8~nm.}
        \label{fig:HS_CLs_15keV}
    \end{figure}

\cleardoublepage

\section{\label{sec:level1}References}
\bibliography{references.bib}
\bibliographystyle{apsrev4-1}